\documentclass[aps,pre,groupedaddress,twocolumn]{revtex4}
\pdfoutput=1

\usepackage{hyperref}
\usepackage{subfigure}
\usepackage{color}
\usepackage{graphicx}
\usepackage{natbib}
\usepackage{doi}

\begin{document}
\bibliographystyle{plainnat}
\expandafter\ifx\csname urlprefix\endcsname\relax\def\urlprefix{URL }\fi

\DeclareGraphicsExtensions{.pdf, .jpg}

\title{\large
 Physical vacuum is a special superfluid medium}

\large
\author{Valeriy I. Sbitnev}
\email{valery.sbitnev@gmail.com}
\address{St. Petersburg B. P. Konstantinov Nuclear Physics Institute, NRC Kurchatov Institute, Gatchina, Leningrad district, 188350, Russia;\\
 Department of Electrical Engineering and Computer Sciences, University of California, Berkeley, Berkeley, CA 94720, USA
}


\date{\today}

\begin{abstract}
The Navier-Stokes equation contains two terms which have been subjected to slight modification:  (a) the viscosity term depends of time (the viscosity in average on time is zero, but its variance is non-zero);  (b) the pressure gradient contains an added term describing the quantum entropy gradient multiplied by the pressure.  Owing to these modifications, the Navier-Stokes equation can be reduced to the Schr\"odinger equation describing  behavior of a particle into the vacuum being as a superfluid medium. Vortex structures arising in this medium show infinitely long life owing to zeroth average viscosity. The non-zero variance describes exchange of the vortex energy with zero-point energy of the vacuum. Radius of the vortex  trembles around some average value. This observation sheds the light to the Zitterbewegung phenomenon. The long-lived vortex has a non-zero core where the vortex velocity vanishes. 
\\

\end{abstract}

\maketitle

\large

\section{\label{sec:level1}Introduction}

 A dramatic situation in physical understanding of the nature emerged in the late of 19th century. 
 Observed phenomena on micro scales came into contradiction with the general positions of classical physics. 
 It was a time of the origination of new physical ideas explaining these phenomena.
 Actually, in a very short period, postulates of the new science, quantum mechanics were formulated.
 The Copenhagen interpretation was first who proposed an ontological basis of quantum mechanics~\citep{Hartle2005}.
 These positions can be stated in the following points: (a) the description of nature is essentially probabilistic;
 (b) a quantum system is completely described by a wave function;
 (c) the system manifests wave--particle duality;
 (d) it is not possible to measure all variables of the system at the same time;
 (e) each measurement of the quantum system entails the collapse of the wave function.
 
 Can one imagine a passage of a quantum particle (the heavy fullerene molecule~\citep{JuffmannEtAl2009}, for example)
 through all slits, in once, at the interference experiment? Following the Copenhagen interpretation,
 the particle does not exists until it is registered. Instead, 
 the wave function represents it existence within an experimental scene~\citep{BruknerZeilinger2005}. 

 Another interpretation was proposed by Louis de Broglie~\citep{deBroglie1987}, which permits to explain such an experiment.
 In de Broglie's wave mechanics and the double solution theory there are two waves. There is the wave function that is a mathematical
 construct. It does not physically exist and is used to determine the probabilistic results of experiments. There is also a physical
 wave guiding the particle from its creation to detection. As the particle moves from a source to a detector, the particle perturbs
 the wave field and gets a reverse effect from it. As a result, the physical wave guides the particle along some optimal trajectory,  Bohmian trajectory~\cite{Bohm1952a, Bohm1952b},
 up to its detection.

 A question arises, what is the de Broglie physical wave? Recently, Couder and Fort~\citep{CouderForte2006} has executed
 the experiment with the classical oil droplets bouncing on the oil surface. A remarkable observation is that an ensemble of the droplets
 passing through the barrier having two gates shows the interference fringes typical for the two slit experiment. Their explanation
 is that the droplet while moving on the surface induces on this surface the weak Faraday waves. The latter provide the guidance conditions
 for the droplets. In this perspective, we can draw conclusion that the de Broglie physical wave can be represented by perturbations of
 the ether when the particle moves through it. In order to describe behavior of such an unusual medium we shall use the Navier-Stokes equation
 with slightly modified some terms. As the final result we shall get the Schr\"odinger equation.
.
 In physical science of the New time the assumption for the existence of the ether medium was originally used to explain propagation of light
 and the long-range interactions. As for the propagation of light, the wave ideas of Huygens and Fresnel require the existence of a continuous
 intermediate environment between a source and a receiver of the light - the light-bearing ether. It is instructive to compare here the two
 opposite doctrines about the nature of light belonging to Sir Isaac Newton and Christian Huygens. Newton maintained the theory that the light
 was made up of tiny particles, corpuscles. They spread through an empty space in accordance with the law of the classical mechanics.
 Christian Huygens (a contemporary of Newton), believed that the light was made up of waves vibrating up and down perpendicularly to
 the direction of its propagation, as waves on a water surface. One can imagine all space populated everywhere densely by Huygens's vibrators.
 All vibrators are silent until a wave reaches them. As soon as a wave front reaches them, the vibrators begin to radiate waves on
 the frequency of the incident wave. So, the infinitesimal volume $\delta V$  is populated by infinite amount of the vibrators with frequencies of
 visible light. These vibrators populate the ether facilitating propagation of the light waves through the space.

 In order to come to idea about existence of the intermediate medium (ether) that penetrates overall material world, we begin from the
 fundamental laws of classical physics. Three Newton's laws first published in Mathematical Principles of Natural Philosophy
 in 1687~\citep{Motte1846} we recognize as basic laws of physics. Namely: (a) the first law postulates existence of inertial reference frames:
 an object that is at rest will stay at rest unless an external force acts upon it; an object that is in motion will not change its velocity
 unless an external force acts upon it. The inertia is a property of the bodies to resist to changing their velocity; 
 (b)~the second law states: the net force applied to a body with a mass M is equal to the rate of change of its linear momentum in
 an inertial reference frame
\begin{equation}
\label{eq=1}
  {\vec F} = M{\vec a}= M{{d{\vec {\mathit v}}}\over{d\,t}};
\end{equation}
 (c) the third law states: for every action there is an equal and opposite reaction.

 Leonard Euler had generalized Newton's laws of motion on deformable bodies that are
 assumed as a continuum~\citep{Frisch2008}. We rewrite the second Newton's law for
 the case of deformed medium. Let us imagine that a volume ${\Delta V}$ contains
 a fluid medium of the mass M. We divide Eq.~(\ref{eq=1}) by ${\Delta V}$ and determine
 the time-dependent mass density $\rho_{_{M}} = M/{\Delta V}$~\citep{Sbitnev2014}.
 In this case ${\vec F}$ is understood as the force per volume.
 Then the second law in this case takes a form:
\begin{equation}
\label{eq=2}
  {\vec F} = {{d\rho_{_{M}}{\vec {\mathit v}}}\over{d\,t}} =
  \rho_{_{M}}{{d{\vec {\mathit v}}}\over{d\,t}}+{\vec {\mathit v}}{{d\rho_{_{M}}}\over{d\,t}}.
\end{equation}
 The total derivatives in the right side can be written down through partial derivatives:
\begin{eqnarray}
\label{eq=3}
 {{d\rho_{_{M}}}\over{d\,t}}   &=& {{\partial\rho_{_{M}}}\over{\partial\,t}}
  + ({\vec {\mathit v}}\,\nabla)\rho_{_{M}},\\
 {{d{\vec {\mathit v}}}\over{d\,t}}\; &=&\; {{\partial{\vec {\mathit v}}}\over{\partial\,t}}
  \;+\; ({\vec {\mathit v}}\,\nabla){\vec {\mathit v}}.
\label{eq=4}
\end{eqnarray}
 Eq.~(\ref{eq=3}) equated to zero is seen to be the continuity equation.
 As for Eq.~(\ref{eq=4}) we may rewrite the rightmost term 
 in detail
\begin{equation}
\label{eq=5}
 ({\vec {\mathit v}}\,\nabla){\vec {\mathit v}} = \nabla\;{{{\mathit v}^2}\over{2}}
 -[{\vec {\mathit v}}\times[\nabla\times{\vec {\mathit v}}]].
\end{equation}
 As follows from this formula the first term, multiplied by the mass, is gradient of
 the kinetic energy. It is a force applied to the fluid element for its shifting on the
 unit of length,~$\delta s$. The second term is acceleration of the fluid element
 directed perpendicularly to the velocity. Let the fluid element move along some curve in
 3D space. Tangent to the curve in each point refers to orientation of the body motion.
 In turn, the vector ${\vec\omega}=[\nabla\times{\vec {\mathit v}}]$ is orientated perpendicularly
 to the plane, where an arbitrarily small segment of the curve lies.
 This vector characterizes a quantitative measure of the
 vortex motion and it is called {\it vorticity}. 
 Vector product $[{\vec {\mathit v}}\times{\vec\omega}]$ is perpendicular to the both vectors ${\vec {\mathit v}}$  and ${\vec\omega}$.
 It represents the Coriolis acceleration of the body
 under rotating it around the vector  ${\vec\omega}$.

 The term~(\ref{eq=5}) entering in the Navier-Stokes equation~\citep{LandauLifshitz1987, KunduCohen2002}
 is responsible for emergence of vortex structures. The Navier-Stokes equation stems from Eq.~(\ref{eq=2}) if we omit the rightmost term, 
 representing the continuity equation, and specify forces in this equation in detail~\citep{Sbitnev2013b}:
\begin{eqnarray}
\nonumber &&
 \rho_{_{M}}\Biggl(
 {{\partial {\vec {\mathit v}}}\over{\partial\,t}}
 + ({\vec {\mathit v}}\cdot\nabla){\vec {\mathit v}}
       \Biggr) 
\\
  &=& {{{\vec F}}\over{\Delta V}}
  + \mu(t)\,\nabla^{\,2}{\vec {\mathit v}} - \rho_{_{M}}\nabla \Biggl( {{P}\over{\rho_{_{M}}}} \Biggr).
\label{eq=6}
\end{eqnarray}
 This equation contains two modifications represented in the two last terms from the right side: the dynamic viscosity $\mu$ depends on
 time and the rightmost term has a slightly modified view, namely 
 $\rho_{_{M}}\nabla(P/\rho_{_{M}})=\nabla P - P \nabla\ln(\rho_{_{M}})$.
 This modification will be important for us when we shall begin to derive the Schr\"odinger equation.
 However first we shall examine the Helmholtz vortices with time dependent viscosity.

\section{\label{sec:level2}Vortex dynamics}

 The second term from the right in Eq.~(\ref{eq=2}) represents the viscosity of the fluid ($\mu$  is the dynamic viscosity, its units are
 N$\cdot$s/m$^2$ = kg/(m$\cdot$s)). Let us suppose that the fluid is ideal, barotropic, and the mass forces are conservative~\cite{KunduCohen2002}.
 At assuming that the external force is conservative, we apply to this equation the operator curl. We get right away the equation for
 the vorticity:
\begin{equation}
\label{eq=7}
 {{\partial\,{\vec\omega}}\over{\partial\,t}}
 + ({\vec\omega}\cdot\nabla){\vec {\mathit v}}
 = \nu\nabla^{\,2}{\vec\omega}.
\end{equation}
 Here $\nu=\mu/\rho_{_{M}}$ is the kinematic viscosity. Its dimension is [m$^2$/s].
 It corresponds to the diffusion coefficient. For that reason the energy stored in the
 vortex will dissipate in thermal energy. As a result, the vortex with the lapse of time
 will disappear.

 With omitted the term from the right (i.e., $\nu=0$) the Helmholtz theorem reads: (i)~if fluid particles form, in any moment of the time,
 a vortex line, then the same particles support the vortex line both in the past and in the future; (ii)~ensemble of the vortex lines traced
 through a closed contour forms a vortex tube. Intensity of the vortex tube is constant along its length and does not change in time.
 The vortex tube (a)~either goes to infinity by both endings; (b)~or these endings lean on walls of bath containing the fluid; 
 (c)~or these endings are locked to each on other forming a vortex ring. 

 Assuming that the fluid is a physical vacuum, which meets the requirements specified earlier, we must say that the viscosity vanishes.
 In that case, the vorticity $\vec\omega$  is concentrated in the center of the vortex, i.e., in the point. Mathematical representation of the vorticity
 is  $\delta$-function. Such singularity can be a source of possible divergences of computations in further. 

 We shall not remove the viscosity. Instead of that, we hypothesize that even if there is an arbitrary small viscosity, because of
 the zero-point oscillations in the vacuum, the vortex does not disappear completely. The vortex can be a long-lived object.
 The foundation for that hypothesis is the observation (performed
 by French scientific team~\cite{CouderForte2006, CouderEtAll2005, ProtiereEtAll2006, EddiEtAll2011})
 of behavior of the droplets moving on the oil surface, on which the waves Faraday exist. Here an important moment is that the Faraday waves
 are supported slightly below the super-critical threshold. Due to this trick the droplets can live on the oil surface arbitrary long,
 before they disappear in the oil. The Faraday waves that are supported near the super-critical threshold may play a role analogous to
 the zero-point oscillations of the vacuum. 

 Observe that the bouncing droplet simulates some aspects of quantum mechanics, stimulating theoretical investigations
 in this area~\citep{Grossing2009, Grossing2010, GrossingEtAl2011, Grossing2013, HarrisBush2014, HarrisEtAl2014, BradyAnderson2014, Vervoort2014}. 
 It is interesting to note in this place that Gr\"ossing considers a quantum particle as a dissipative phase-locked steady state,
 where an amount of zero-point energy of the wave-like environment is absorbed by the particle, and then during a characteristic relaxation
 time is dissipated into the environment again~\citep{Grossing2010}.

 Here we shall give a simple model of such a picture. Let us look on the vortex tube in its cross-section which is oriented along
 the axis $z$ and its center is placed in the coordinate origin of the plane $(x, y)$.
 Eq. (\ref{eq=7}), written down in the cross-section of the vortex, is as follows
\begin{equation}
\label{eq=8}
 {{\partial\omega}\over{\partial\,t}} =
 \nu g(t)\Biggl(
    {{\partial^{\,2}\omega}\over{\partial r^2}}
  + {{1}\over{r}}{{\partial\omega}\over{\partial r}}
    \Biggr).
\end{equation}
 Here we do not write a sign of vector on the top of $\omega$ since $\omega$ is oriented
 strictly along the axis $z$. We introduce time-dependent the kinematic viscosity. For the sake of simplicity,
 let it be looked as
\begin{equation}
\label{eq=9}
  g(t) = \cos(\Omega t+\phi) = {{e^{{\bf i}(\Omega t+\phi)} + e^{-{\bf i}(\Omega t+\phi)}}\over{2}}
\end{equation}
 where $\Omega$  is an oscillation frequency and  $\phi$ is the uncertain phase.  

 Solution of the equation~(\ref{eq=8}) in this case is as follows
\begin{widetext}
\begin{equation}
\label{eq=10}
 \omega(r,t) ={{\mit\Gamma}\over{4\pi(\nu/\Omega)(\sin(\Omega t+\phi)+n))}}
 \exp\Biggl\{
   - {{r^2}\over{4\pi(\nu/\Omega)(\sin(\Omega t+\phi)+n))}}
  \Biggr\}.
\end{equation}
 Here $\mit\Gamma$  is the integration constant having dimension m$^2$/s.
 An extra number $n > 1$. It prevents appearance of singularity in the cases when $\sin(\Omega t+\phi)$  tends to $-1$.
 This function at choosing the parameters ${\mit\Gamma}=1$, $\nu=1$ , $\Omega=\pi$  and $n = 16$ is shown in Fig.~\ref{fig=1}(a).  

 The velocity of the fluid matter around the vortex results from the integration of the vorticity function
\begin{equation}
\label{eq=11}
 {\mathit{\vec v}} = {{1}\over{r}}\int\limits_{0}^{r}\omega(r',t)r'dr' =
 {{\mit\Gamma}\over{2\pi r}}\Biggl(1 - 
 \exp\Biggl\{
   - {{r^2}\over{4\pi(\nu/\Omega)(\sin(\Omega t+\phi)+n))}}
  \Biggr\}
  \Biggr).
\end{equation}
 Fig.~\ref{fig=1}(b) shows behavior of this function at the same input parameters. 
\begin{figure*}[htb!]
  \centering
  \begin{picture}(200,200)(130,80)
      \includegraphics[scale=0.55]{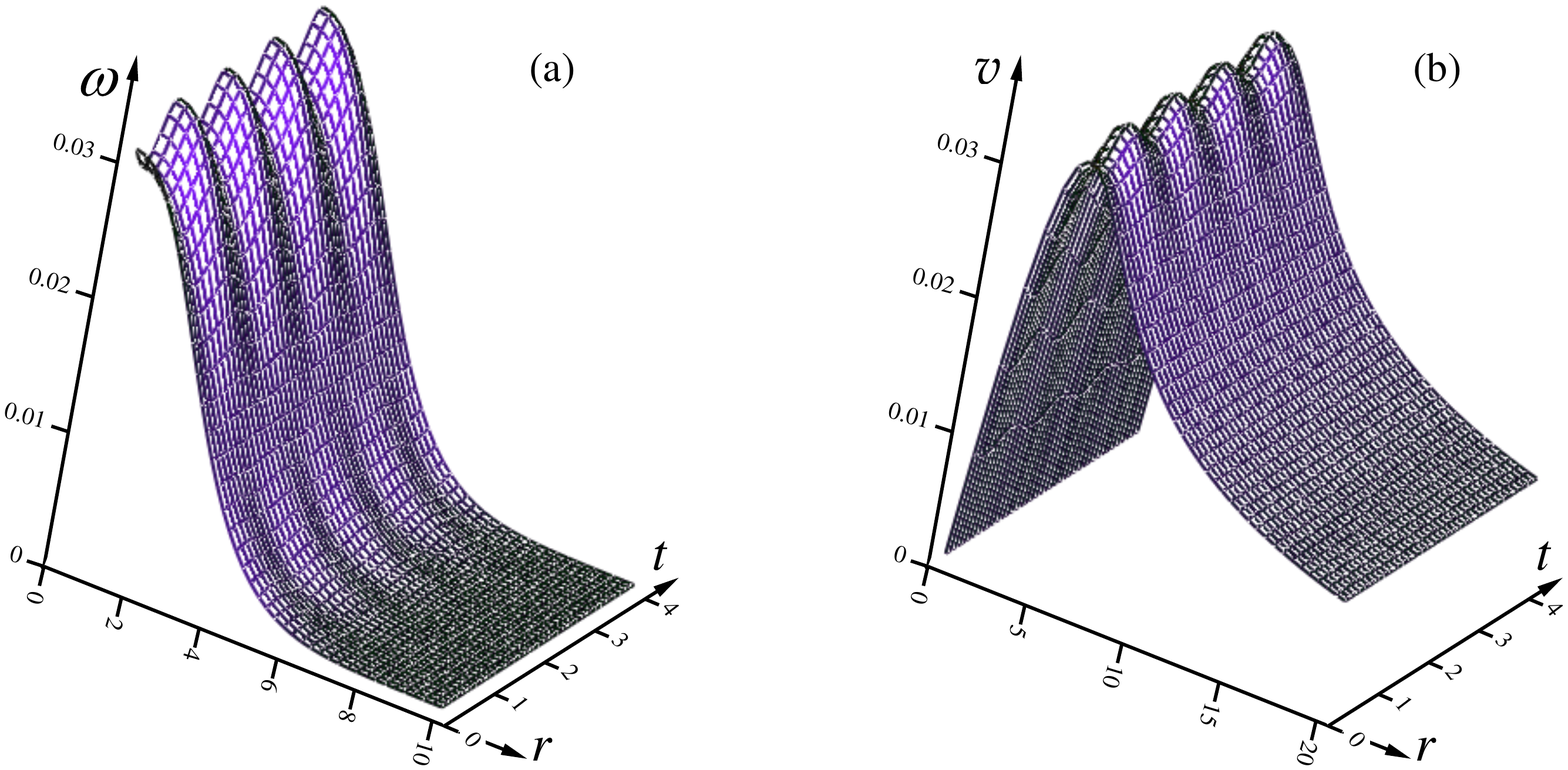}
  \end{picture}
  \caption{
 Vorticity $\omega(r,t)$  and velocity ${\mathit v}(r,t)$ as functions of $r$ and $t$
 for ${\mit\Gamma}=1$,  $\nu=1$, $\Omega=\pi$,  and $n=16$. These parameters are conditional in order to show clearly oscillations
 of the vortex in time. The solution does not decay with time.
  }
  \label{fig=1}
\end{figure*}

 In particular, for $n = 0$ and $\Omega t \ll 1$  this solution is close to the Lamb-Oseen vortex solution~\cite{WuMaZhou2006}
\begin{equation}
\label{eq=12}
  \omega(r,t)={{\mit\Gamma}\over{4\pi\nu t}}e^{-r^2/4\nu t},
  \hspace{64pt}
  {\mathit v}(r,t)= {{\mit\Gamma}\over{4\pi r}}\biggl( 1 - e^{-r^2/4\nu t} \biggr).
\end{equation}
\end{widetext}
 It is seen that the Lamb-Oseen vortex solution decays with time the faster, the more $\nu$.

 One can see from the solutions~(\ref{eq=10}) and~(\ref{eq=11}), depending on the distance to the center the functions
 ${\omega}(r,t)$   and ${\mathit v}(r,t)$ show typical behavior for the vortices. The both functions do not decay with time,
 however. Instead of that, they demonstrate pulsations on the frequency $\Omega$. Amplitude of the pulsations is the smaller,
 the larger value of the parameter $n$. At $n$ tending to infinity the amplitude of the pulsations tends to zero.
 At the same time the vortex disappears entirely. 

\begin{figure}
  \centering
  \begin{picture}(200,80)(20,25)
      \includegraphics[scale=0.7]{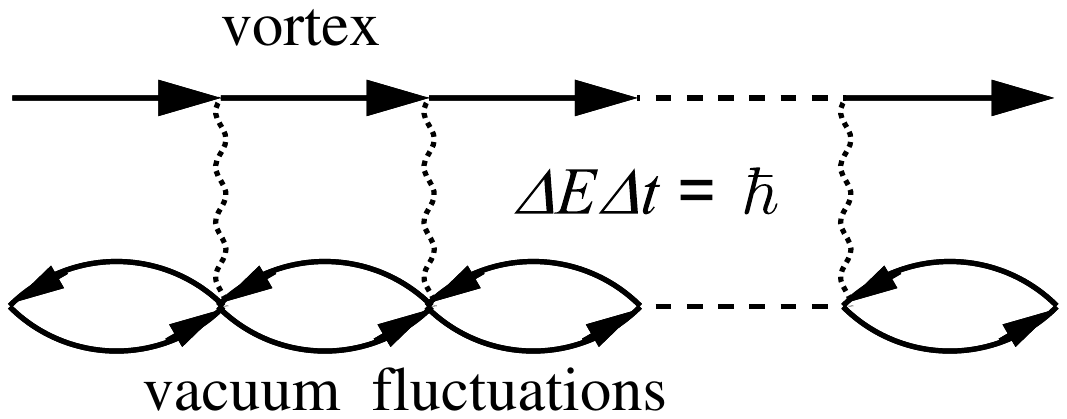}
  \end{picture}
  \caption{
  Periodic energy exchange between the vortex and vacuum fluctuations  }
  \label{fig=2}
\end{figure}
\begin{figure}
  \centering
  \begin{picture}(200,220)(35,5)
      \includegraphics[scale=0.65]{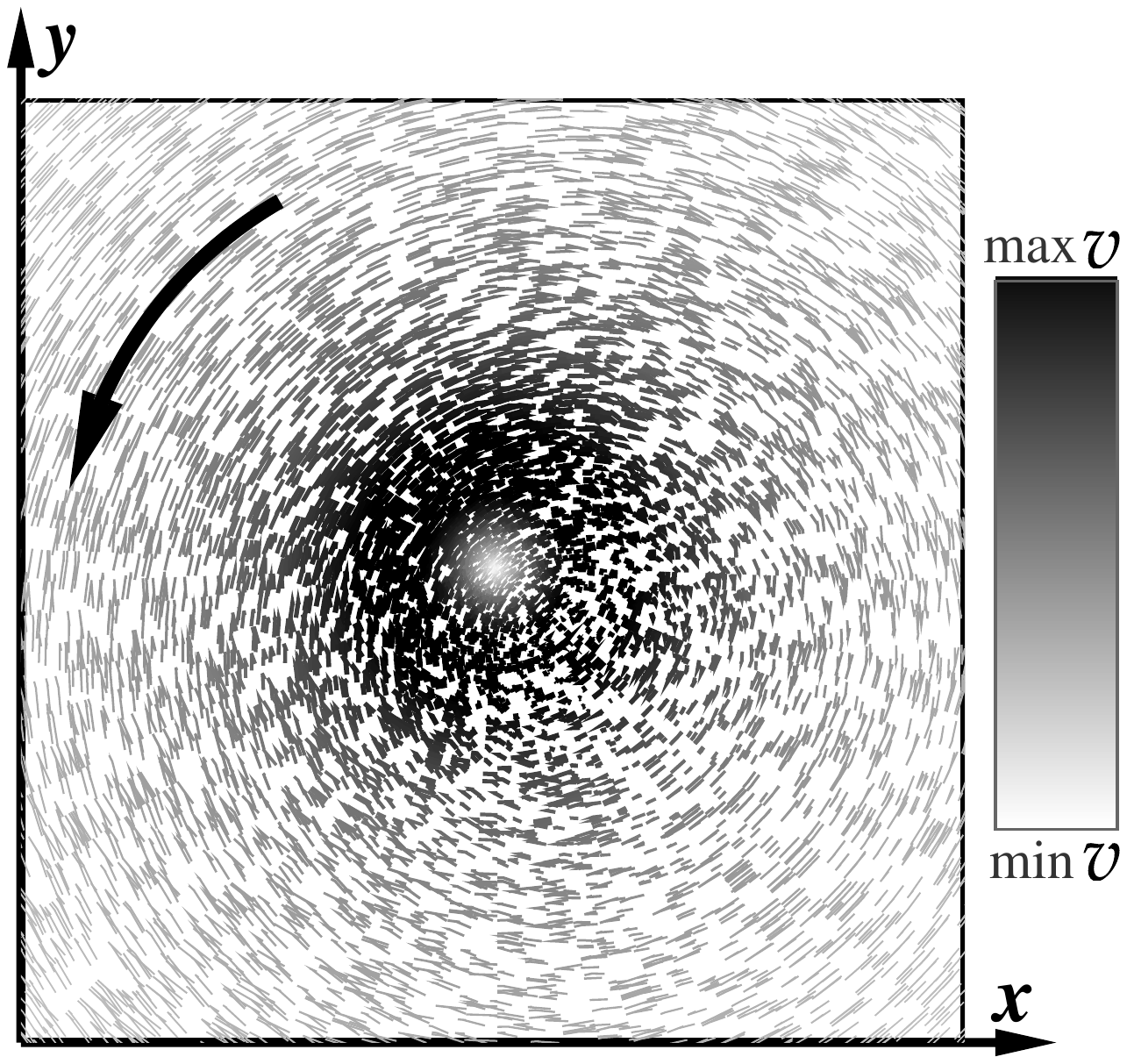}
  \end{picture}
  \caption{
  Cross-section of the vortex tube in the plane $(x, y)$. Values of the velocity $\mathit v$ are shown in grey ranging from light grey
 ($\min\mathit v$) to dark grey ($\max\mathit v$).  Density of the pixels represents magnitude of the vorticity .
 Core of the vortex $\omega$ is well visible in the center.
  }
  \label{fig=3}
\end{figure}
 The undamped solution was obtained thanks to assumption, that the kinematic viscosity is a periodic function of time,
 namely, $\nu g(t)=\nu\cos(\Omega t+\phi)$. The viscosity in the quantum realm is not a good concept, however. Most likely, it manifests
 itself through interaction of the quantum object with vacuum fluctuations. According to Eq.~(\ref{eq=9}), there are half-periods
 when the energy of the vortex is lost at scattering on the vacuum fluctuations, and there are other half-periods when the vacuum
 returns this energy to the vortex, Fig.~\ref{fig=2}. On the whole, the viscosity of the fluid medium, within which the vortex tube evolves,
 in the average remains at zero. It can mean that this medium is superfluid. Such a scenario is not unusual. For example, at transition
 of helium to the superfluid phase~\cite{Volovik2003} coherent Cooper pairs of electrons arise through the exchange by phonons.
 This attraction is due to the electron-phonon interaction. The phonons are thermal excitations of a lattice. In that case,
 they play a role of the background medium.

 Qualitative view of the vortex tube in its cross-section is shown in Fig.~\ref{fig=3}.
 Values of the velocity $\mathit v$ are shown by grey color ranging from light grey (minimal velocities) to dark grey (maximal ones).
 A visual image of this picture can be a hurricane (tropical cyclone~\citep{TropicalCyclne}) shown from the top. In the center of the vortex, a so-called eye
 of the hurricane (the vortex core) is well viewed. Here it looks as a small light grey disk, where the velocities have small values.
 In the very center of the disk, in particular, the velocity vanishes. Observe that in the region of the hurricane eye a wind is really
 very weak, especially near the center. This is in stark contrast to conditions in the region of the eyewall, where the strongest winds
 exist (in Fig.~\ref{fig=3} it looks as a dark grey annular region enclosing the light grey inner area). The eyewall of the vortex tube
 (a zone where the velocity reaches maximal values) has the nonzero radius. 

  Let us find the radius of the vortex core. In order to evaluate this radius we equate to zero the first derivative by
 $r$ of equation~(\ref{eq=11})
\begin{eqnarray}
\nonumber
&&
  \exp\Biggl\{
       {{r^2}\over{4\pi(\nu/\Omega)(\sin(\Omega t+\phi)+n))}}
      \Biggr\}
\\
\label{eq=13}
 && =
 2        {{r^2}\over{4\pi(\nu/\Omega)(\sin(\Omega t+\phi)+n))}} + 1.
\end{eqnarray}
 The radius is a root of this equation
\begin{equation}
\label{eq=14}
   r_{v}=2\sqrt{a_{0}(n+\sin(\Omega t + \phi))}\cdot\sqrt{{\nu}\over{\Omega}}.
\end{equation}
 Here $a_0\approx1.2564312$ is a root of the equation $\ln(2a_0 + 1) - a_0 = 0$.
 One can see, that $r_v$ is an oscillating function.
 The larger $\Omega$, the more quickly the vortex trembles.
 As $\Omega$ increases, the vortex radius decreases. However it grows with increasing the number $n$.
 Let us evaluate the radius $r_v$ at choosing the viscosity $\nu$ equal to ${\bar\nu}=\hbar/2m$. Here $\bar\nu$ is the diffusion coefficient of the Brownian sub-quantum particles wandering in the Nelson's aether~\citep{Nelson1966}, see {\bf Appendix}. In the case of electron ${\bar\nu} \approx 5.79\cdot10^{-5}$  m$^2$/s. As for $\Omega$, let it be equal to $2mc^2/\hbar$, or approximately $1.6\cdot10^{\,21}$ radians per second for electron. Here $c$ is the speed of light
and $m$ is the electron mass. Then we have $(\nu/\Omega)^{1/2} \approx 1.93\cdot10^{-13}$~m. This length is seen to be smaller then the Compton wavelength, $\lambda_{_{C}}= 2.426\cdot10^{-12}$~m, in about 12 times. So, for choosing $n \approx 31$ we find from~(\ref{eq=14}) that the radius of the vortex is about the Compton wavelength. This number is quite enough to prevent any fatal catastrophe of the vortex. That is, the vortex is a long-lived robust object.
 From the above we see that, on a distance about the Compton wavelength, virtual particles can be involved into a vorticity dancing around the electron core, by polarizing the electron charge. This dancing happens at trembling motion of the electron with the frequency $\Omega=2mc^2/\hbar$. That oscillating motion has a deep relation to the so-called  "Zitterbewegung"~\citep{BarutBracken1981, Hestenes1990}.

 One can give a general solution of Eq.~(\ref{eq=8}) which has the following presentation
\begin{widetext}
\begin{equation}
\label{eq=15}
 \omega(r,t)={{\mit\Gamma}\over{4\pi\Biggl(\int\limits_{0}^{t}\nu(\tau)d\tau}+\sigma^2\Biggr)}
  \exp
  \matrix{
  \left\{\displaystyle
      -{{r^2}\over{4\pi\Biggl(\int\limits_{0}^{t}\nu(\tau)d\tau}+\sigma^2\Biggr)}  
      \right\}},
\end{equation}
\begin{equation}
\label{eq=16}
 {\mathit {v}}(r,t)={{\mit\Gamma}\over{2\pi r}}
 \matrix{
             \left( 1 -
  \exp
  \left\{\displaystyle
      -{{r^2}\over{4\pi\Biggl(\int\limits_{0}^{t}\nu(\tau)d\tau}+\sigma^2\Biggr)}  
      \right\}
           \right)
          }
\end{equation}
\end{widetext}
 The viscosity function $\nu(t)$  is a quasi-periodic function or even is represented by a color noise. The integral of the viscosity
 function memorizes integrally character of the viscosity of the medium. Due to this memory effect, the vortex may live a long enough.
 As for interpretation of these solutions with the quantum-mechanical point of view, we may say that there exists a regular exchange by quanta
 with the vacuum fluctuations, Fig.~\ref{fig=2}. The integral accumulates all cases of the exchange with the vacuum. The constant $\sigma$ 
 having dimension of length, prevents appearance of singularities. One can see that even at $\nu = 0$, but $\sigma>0$, abundance of long-lived
 vortices can exist in the vacuum. Such vortices are "ghosts" in the superfluid being invisible without interaction.

 The viscosity function $\nu(t)$  is a quasi-periodic function or even is represented by a color noise. The integral of the viscosity
 function memorizes integrally character of the viscosity of the medium. Due to this memory effect, the vortex may live a long enough.
 As for interpretation of these solutions with the quantum-mechanical point of view, we may say that there exists a regular exchange by
 quanta with the vacuum fluctuations, Fig.~\ref{eq=2}. The integral accumulates all cases of the exchange with the vacuum.
 The constant $\sigma$  having dimension of length, prevents appearance of singularities. One can see that even at $\nu=0$,
 but $\sigma>0$, abundance of long-lived vortices can exist in the vacuum. Such vortices are "ghosts"
 in the superfluid being invisible without interaction.

\subsection{\label{subsec:level2A}Vortex rings and vortex balls}

\begin{figure}
  \centering
  \begin{picture}(200,120)(-10,25)
      \includegraphics[scale=0.35]{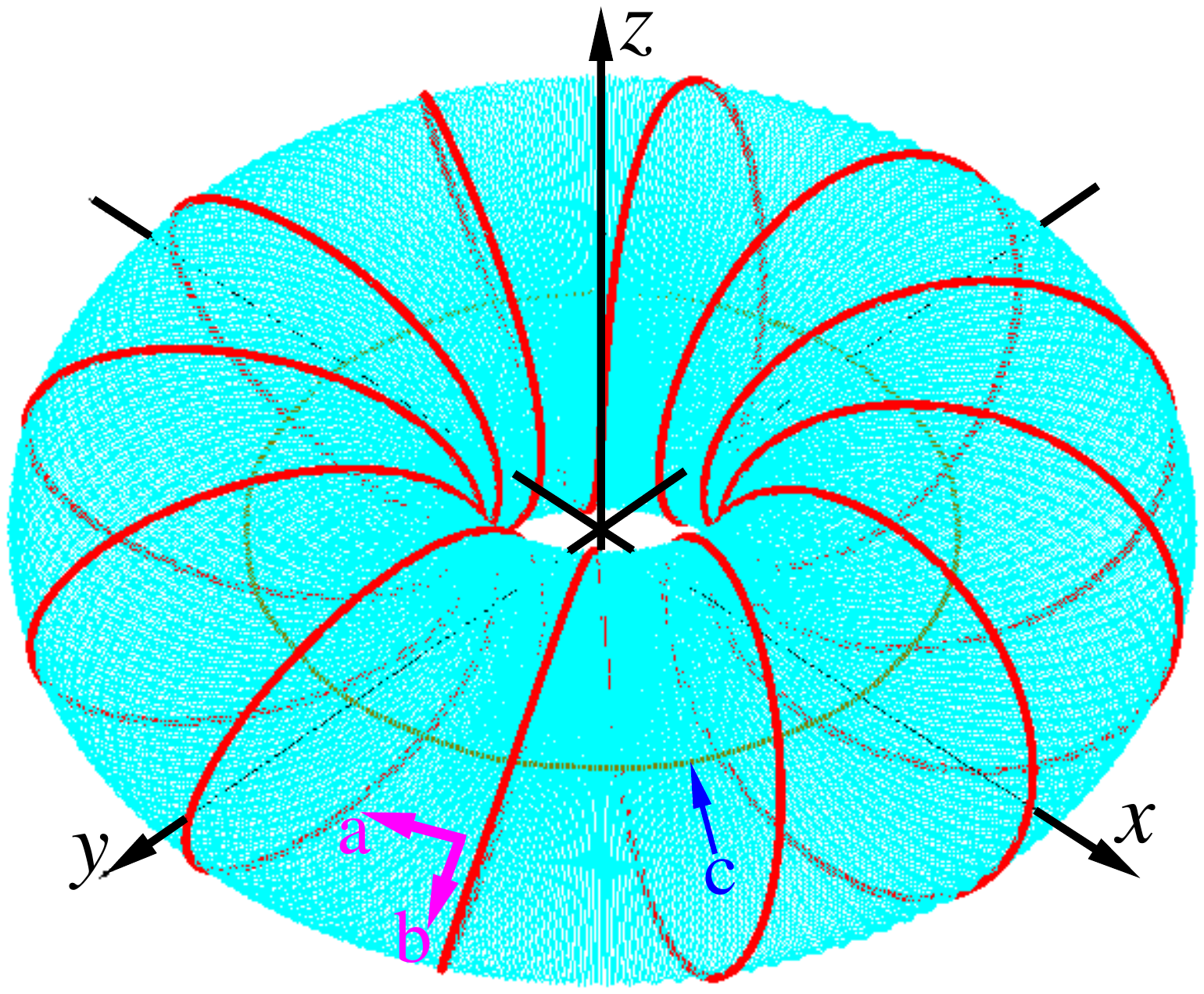}
  \end{picture}
  \caption{
  Helicoidal vortex ring: $r_{0}=2$, $r_{1}=3$, $\omega_{2}=12\omega_{1}$, $\phi_{2}=\phi_{1}=0$.  }
  \label{fig=4}
\end{figure}

 If we roll up the vortex tube in a ring and glue together its opposite ends we obtain a vortex ring. A result of such an operation put into
 the $(x, y)$ plane is shown in Fig~\ref{fig=4}. Position of points on the helicoidal vortex ring~\citep{Sonin2012}
 in the Cartesian coordinate system is given by
\begin{equation}
\label{eq=17}
\hspace{-10pt}
\left\{
\matrix{
  x=(r_{1}+r_{0}\cos(\omega_{2}t+\phi_{2}))\cos(\omega_{1}t+\phi_{1}),\cr
  y=(r_{1}+r_{0}\cos(\omega_{2}t+\phi_{2}))\sin(\omega_{1}t+\phi_{1}),\cr
  z= r_{0}\sin(\omega_{2}t+\phi_{2}).\hspace{104pt}\cr
       }
\right.
\end{equation}
 Here $r_{0}$ is the radius of the tube. And $r_{1}$ represents the distance from the center of the tube (pointed in the figure by arrow c)
 to the center of the torus located in the origin of coordinates $(x,y,z)$. A body of the tube, for the sake of visualization is colored in
 cyan. Eq.~(\ref{eq=17}) parametrized by $t$ gives a helicoidal vortex ring shown in this figure.
 Parameters $\omega_{1}$ and $\omega_{2}$ are
 frequencies of rotation along the arrow a about the center of the torus (about the axis $z$) and rotation along the arrow b about the center
 of the tube (about the axis pointed by arrow c), respectively. Phases $\phi_{1}$ and $\phi_{2}$ have uncertain quantities ranging from 0 to
 $2\pi$. By choosing the phases within this interval with a small increment, we may fill the torus by the helicoidal vortices everywhere
 densely. The vorticity is maximal along the center of the tube. Whereas the velocity of rotation about this center in the vicinity of it
 is minimal. However, the velocity grows as a distance from the center increases. After reaching of some maximal value the velocity further
 begins to decrease. 

\begin{figure}
  \centering
  \begin{picture}(200,150)(-10,10)
      \includegraphics[scale=0.3]{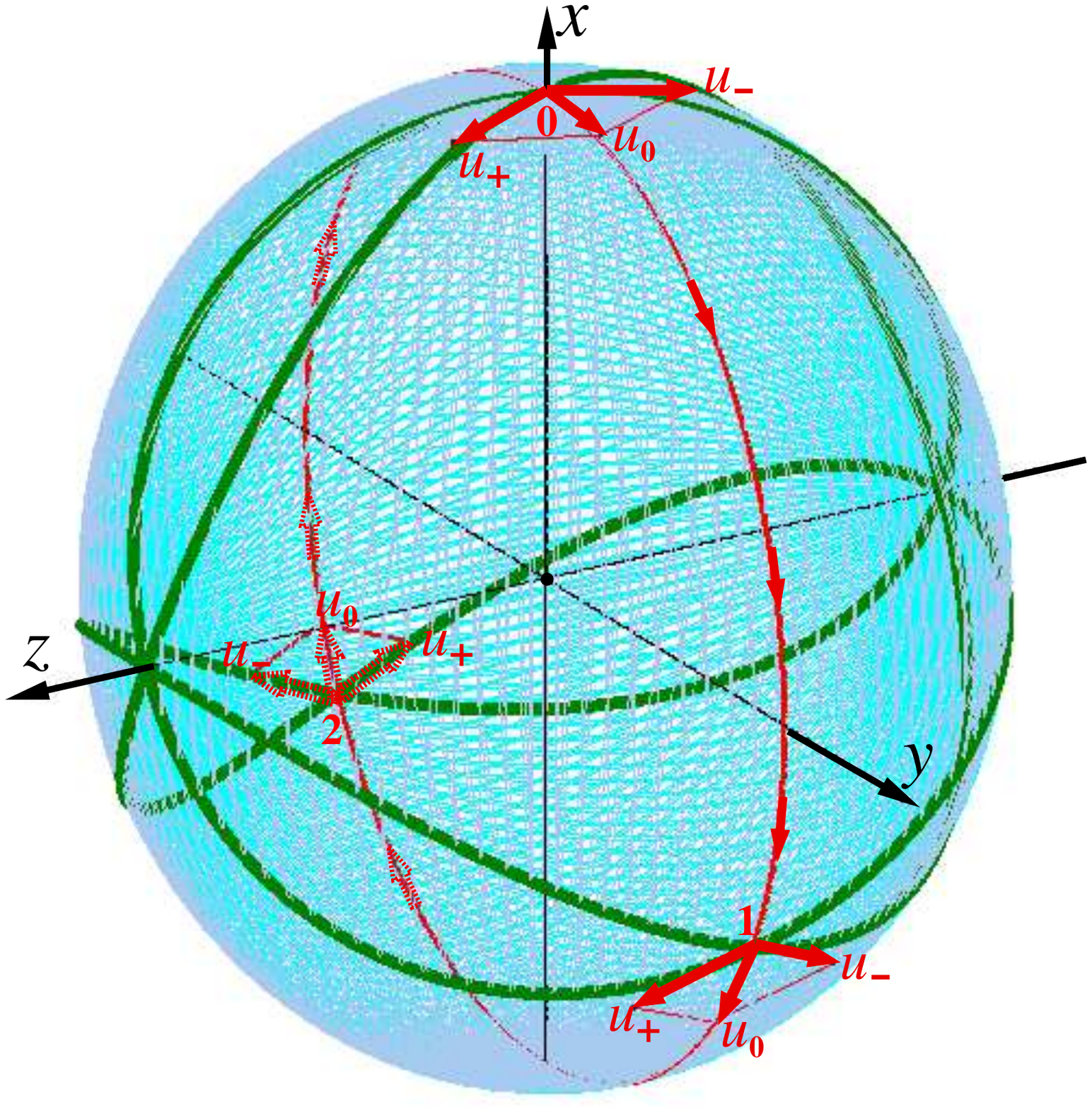}
  \end{picture}
  \caption{
  Helicoidal vortex ring (colored in deep green) convoluted into the vortex ball. 
  The input parameters of the ball are as follows $r_0 = 4$, $r_1= 0.01 \ll 1$, $\omega_2=3\omega_1$, $\phi_2=\phi_1=0$.
  The radius $r_0$ represents a mean radius of the ball, where the velocity ${\mathit v}_0$ reaches a maximal value.
 }
  \label{fig=5}
\end{figure}

 Let the radius $r_{1}$ in Eq.~(\ref{eq=17}) tends to zero. The helicoidal vortex ring in this case will transform into a vortex ring
 enveloping a spherical ball. The vortex ring for the case $r_{0}=4$, $r_{1}\approx 0$, $\omega_{2}=3\omega_{1}$ and $\phi_{2}=\phi_{1}=0$
 drawn by thick curve colored in deep green is shown in Fig.~\ref{fig=5}. Motion of an elementary vortex clot along the vortex ring (along the thick
 curve colored in deep green) takes place with a velocity
\begin{widetext}
\begin{equation}
 {\vec{\mathit v}}_{_{R}} =
 \left\{
 \matrix{
   {\mathit v}_{_{R,x}} =
        - r_{0}\omega_{\,0}\sin(\omega_{\,0}t+\phi_{\,0})\cos(\omega_{1}t+\phi_{1})
        - r_{0}\omega_{1}\cos(\omega_{\,0}t+\phi_{\,0})\sin(\omega_{1}t+\phi_{1})\cr
\hspace{-214pt}         - r_{1}\omega_{1}\sin(\omega_{1}t+\phi_{1}),\cr
   {\mathit v}_{_{R,y}} =
        - r_{0}\omega_{\,0}\sin(\omega_{\,0}t+\phi_{\,0})\sin(\omega_{1}t+\phi_{1})
        + r_{0}\omega_{1}\cos(\omega_{\,0}t+\phi_{\,0})\cos(\omega_{1}t+\phi_{1})\cr
\hspace{-214pt}         +  r_{1}\omega_{1}\cos(\omega_{1}t+\phi_{1}),\cr
   {\mathit v}_{_{R,z}} =
       ~~r_{0}\omega_{\,0}\cos(\omega_{\,0}t+\phi_{\,0}),\hspace{246pt}\cr
        }
 \right.
\label{eq=18}
\end{equation}
\end{widetext}
 The velocity of the clot at the initial time is  
 ${\mathit v}_{_{R,x}}=0$, ${\mathit v}_{_{R,y}}=r_{0}\omega_{1}$, ${\mathit v}_{_{R,z}}=r_{0}\omega_{2}=3r_{0}\omega_{1}$
 (the initial point $(x, y, z) = (4, 0, 0)$ is on the top of the ball).
 We designate this velocity as ${\vec{\mathit v}}_{+}$. Through the time $t=\pi\omega_{1}$ 
 the elementary vortex clot returns to the top position. The velocity in this case is equal to 
 ${\mathit v}_{_{R,x}}=0$, ${\mathit v}_{_{R,y}}=r_{0}\omega_{2}$, ${\mathit v}_{_{R,z}}=-r_{0}\omega_{2}=3r_{0}\omega_{1}$.
 We designate this velocity as ${\vec{\mathit v}}_{-}$.
 Sum of the two opposite velocities, ${\vec{\mathit v}}_{+}$ and ${\vec{\mathit v}}_{-}$, gives the velocity
 ${\vec{\mathit v}}_{0}=(0,2r_{0}\omega_{1},0)$. During $t=(1+3k)\pi/3\omega_{1}$ and $t=(2+3k)\pi/3\omega_{1}$  $(k = 1, 2, \cdots)$
 the clot travels through the positions 1 and 2 both in the forward and in backward directions, respectively. In the vicinity of  these points
 the velocities ${\vec{\mathit v}}_{+}$ and ${\vec{\mathit v}}_{-}$ yield the resulting velocity ${\vec{\mathit v}}_{0}$ directed along
 the circle lying in the plane $(x, y)$.

\begin{figure}
  \centering
  \begin{picture}(200,120)(-20,10)
      \includegraphics[scale=0.27]{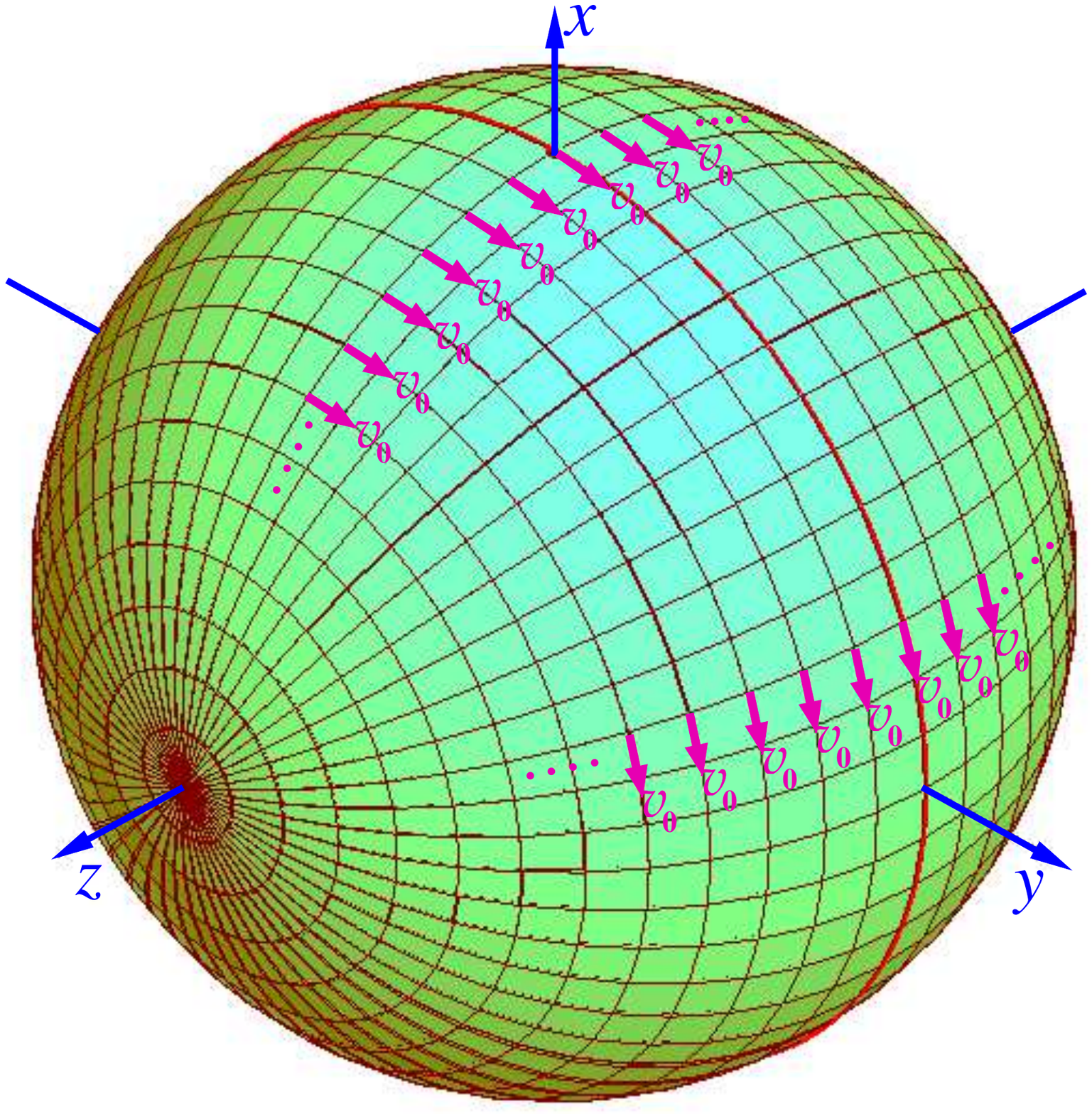}
  \end{picture}
  \caption{
   The vortex ball rotating about axis z with the maximal velocity v0 that is reached on the surface of the ball }
  \label{fig=6}
\end{figure}

 The ball can be filled everywhere densely by other rings at adding them with other phases
 $\phi_{1}$  and $\phi_{2}$ ranging from 0 to $2\pi$  The velocity ${\vec{\mathit v}}_{0}$  for any ring will lie on the same circles centered
 on the axis $z$. We see a dense ball that rolls along the axis $y$, Fig.~\ref{fig=6}. Observe that the ball pulsates on the frequency
 $\Omega$ as it rolls along its path, as it follows from the above computations. Perfect modes describing the rolling ball
 are spherical harmonics~\citep{DorboloEtAl2008}.

\section{\label{sec:level3}Derivation of the Schr\"odinger equation}

 The third term in the right side of Eq.~(\ref{eq=6}) deals with the pressure gradient. One can see, however, it is slightly differ from
 the pressure gradient presented in the customary Navier-Stokes equation~\citep{LandauLifshitz1987, KunduCohen2002}.
 One can rewrite this term in detail
\begin{equation}
\label{eq=19}
  \rho_{_{M}}\nabla\Biggl(
                   {{P}\over{\rho_{_{M}}}}
                   \Biggr)
 = \nabla P - P\nabla\ln(\rho_{_{M}}).
\end{equation}
 The first term, $\nabla P$, is the customary pressure gradient represented in the Navier-Stokes equation. Whereas, 
 the second term,  $P\nabla\ln(\rho_{_{M}})$, is an extra term describing changing the logarithm of the density along increment of length
 (the entropy increment) multiplied by $P$.  It may mean that change of the pressure is induced by change of the entropy per length, or else
 by change of the information flow~\citep{Sbitnev2012, LicataFiscaletti2014} per length. 
 This term has signs typical of the osmotic pressure, mentioned by Nelson~\citep{Nelson1966}.

 Let us consider in this respect the pressure P in more detail. We shall represent the pressure consisting of two parts, $P_1$ and $P_2$.
 We begin from the Fick's law~\citep{Grossing2010}. The law says that the diffusion flux, $\vec J$, is proportional to the negative value of
 the density gradient ${\vec J} =-D\nabla\rho_{_{M}} $. Here $D$ is the diffusion coefficient ${\bar\nu}=\hbar/2m$~\citep{Nelson1966}, 
 see Nelson's definition in {\bf Appendix}. Since the term ${\bar\nu}\nabla{\vec J}$  has dimension of the pressure, we define $P_1$
 as the pressure having diffusion nature
\begin{equation}
\label{eq=20}
  P_1 = {\bar\nu}\nabla{{\vec J}}
 = -{{\hbar^{2}}\over{4m^2}}\nabla^{2}\rho_{_{M}}.
\end{equation}
 Observe that the kinetic energy of the diffusion flux is $(m/2)({\vec J}/\rho_{_{M}})^{2}$ . It means that there exists one more pressure
 as the average momentum transfer per unit area per unit time:
\begin{equation}
\label{eq=21}
  P_2 = {{\rho_{_{M}}}\over{2}}\Biggl(
                                      {{{\vec J}}\over{\rho_{_{M}}}}
                               \Biggr)^2
 = {{\hbar^2}\over{8m^2}}{{(\nabla\rho_{_{M}})^2}\over{\rho_{_{M}}}}.
\end{equation}
 Now we can see that sum of the two pressures, $P_1 + P_2$, divided by $\rho_{_{M}}$  gives a term
\begin{equation}
\label{eq=22}
  Q_{_{M}} = {{\hbar^2}\over{8m^2}}\Biggl(
                                   {{\nabla\rho_{_{M}}}\over{\rho_{_{M}}}} 
                                   \Biggr)^2
 - {{\hbar^2}\over{4m^2}}{{\nabla^2\rho_{_{M}}}\over{\rho_{_{M}}}}.
\end{equation}
 One can see that accurate to the divisor $m$ this term represents the quantum potential. 

 To bring the expression~(\ref{eq=22}) to a form of the quantum potential, we need to introduce instead of the mass density $\rho_{_{M}}$
 the probability density  according to the following presentation
\begin{equation}
\label{eq=23}
   \rho_{_{M}} = {{M}\over{\Delta V}} = {{mN}\over{\Delta V}} = m\rho.
\end{equation}
 Here the mass $M$ is a product of an elementary mass $m$ by the number of these masses,
 $N$, filling the volume $\Delta V$.
 Then the mass density $\rho_{_{M}}$ is defined as a product of
 the elementary mass $m$ by the density of quasi-particles $\rho=N/\Delta V$.
 We may imagine such a set of the quasi-particles as a collection of particles
 which comes  from the realm of classical Lagrange fluids~\citep{JackiwEtAl2004}.

 Let us divide the Navier-Stokes equation Eq.~(\ref{eq=6}) by the probability density $\rho$. We obtain
\begin{eqnarray}
\nonumber &&
 m\Biggl(
 {{\partial {\vec {\mathit v}}}\over{\partial\,t}}
 + ({\vec {\mathit v}}\cdot\nabla){\vec {\mathit v}}
       \Biggr) 
\\
  &=& {{{\vec F}}\over{N}}
  + \nu(t)\,\nabla^{\,2}m{\vec {\mathit v}} - \nabla Q.
\label{eq=24}
\end{eqnarray}
 Here ${\vec F}/N$  is the force per one the quasi-particle. The kinetic viscosity $\nu(t)=\mu(t)/\rho_{_{M}}$ is represented through
 the diffusion coefficient ${\bar\nu}=\hbar/2m$~\citep{Nelson1966}. Namely, $\nu(t) = 2{\bar\nu}g(t)=\nu g(t)$, where $\nu=\hbar/m$ and $g(t)$ is the dimensionless time dependent function.
 The function $Q$ here is the real quantum potential 
\begin{equation}
\label{eq=25}
  Q = {{\hbar^2}\over{8m}}\Biggl(
                                   {{\nabla\rho}\over{\rho}} 
                                   \Biggr)^2
 - {{\hbar^2}\over{4m}}{{\nabla^2\rho}\over{\rho}}.
\end{equation}
 Gr\"ossing noticed that the term  $\nabla Q$, the gradient of the quantum potential, describes a completely thermalized fluctuating force
 field~\citep{Grossing2009, Grossing2010}. Here the fluctuating force is expressed via the gradient of the pressure divided by the density
 distribution of sub-quantum particles chaotically moving in the environment. Perhaps, they are virtual particle-antiparticle pairs.

 Since the pressure provides a basis of the quantum potential, as was shown above, it would be interesting to interpret an osmotic nature
 of the pressure~\citep{Nelson1966}. The interpretation can be the following (see {\bf Appendix}): a semipermeable membrane where the osmotic
 pressure manifests itself is an instant, which divides the past and the future (that is, the 3D brane of our being represents the
 semipermeable membrane in the 4D world). In other words, the thermalized fluctuating force field described by
 Gr\"ossing~\citep{Grossing2009, Grossing2010, GrossingEtAl2011, Grossing2013} is asymmetric with respect to the time arrow.

\subsection{\label{subsec:level3A}Transition to the Schr\"odinger equation}

 The current velocity $\mathit v$ contains two component --  irrotational and solenoidal~\citep{KunduCohen2002} that relate to vortex-free
 and vortex motions of the medium, respectively. The basis for the latter is the Kelvin-Stokes theorem.  Scalar and vector fields underlie
 of manifestation of the irrotational and solenoidal velocities
\begin{equation}
\label{eq=26}
   {\mathit{\vec v}}={\mathit{\vec v}}_{_{S}}+{\mathit{\vec v}}_{_{R}}
   = {{1}\over{m}}\nabla S + {\mathit{\vec v}}_{_{R}}.
\end{equation}
 Here subscripts $S$ and $R$ hint to scalar and vector (rotational) potentials underlying emergence of these two components of the velocity.
 These velocities satisfy the following equations
\begin{equation}
\label{eq=27}
 \left\{
    \matrix{
           (\nabla\cdot{\mathit{\vec v}}_{_{S}}) \ne 0, & [\nabla\times{\mathit{\vec v}}_{_{S}}]=0, \cr
           (\nabla\cdot{\mathit{\vec v}}_{_{R}})  =  0, & [\nabla\times{\mathit{\vec v}}_{_{R}}]={\vec\omega}. \cr
           }
 \right.
\end{equation}
 The scalar field is represented by the scalar function $S$ - action in classical mechanics. Both velocities are perpendicular to each other.
 We may define the momentum and the kinetic energy
\begin{equation}
\label{eq=28}
 \left\{\,
    \matrix{
           {\vec p} = m{\mathit{\vec v}} = \nabla S + m{\mathit{\vec v}}_{_{R}}, \cr\cr
           {\displaystyle
            m{{{\mathit v}^2}\over{2}} = 
           {{1}\over{2m}}(\nabla S)^2 + m {{{\mathit v}_{_{R}}^2}\over{2}}}. \cr
           }
 \right.
\end{equation}
\begin{widetext}
 Now we may rewrite the Navier-Stokes equation~(\ref{eq=24}) in the more detailed form
\begin{eqnarray}
\nonumber
  {{\partial}\over{\partial\,t}}(\nabla S + m{\mathit{\vec v}}_{_{R}})
   &+& 
  \Biggl\{ {{1}\over{2m}}\bigl(    
                      (\nabla S)^2 + m^2{\mathit v}_{_{R}}^2
                      \bigr)
    +   \bigl[{\vec\omega}\times
                (\nabla S + m{\mathit{\vec v}}_{_{R}})
        \bigr]
 \Biggr\}
\\ \nonumber
\\
   &=& - \nabla U - \nabla Q
    + 
          \nu(t)\nabla^2(\nabla S + m{\mathit{\vec v}}_{_{R}})
\label{eq=29}
\end{eqnarray}
 The term embraced by the curly brackets in the first line stems from 
 $({\vec{\mathit v}}\nabla){\vec{\mathit v}}=\nabla{\mathit v}^2/2 + [{\vec\omega}\times{\vec{\mathit v}}]$, see Eq.~(\ref{eq=5}).
As for the external force in the Navier-Stokes equation~(\ref{eq=24}), we take into account that it is conservative,
 i.e.,  ${\vec F}/N = -\nabla U$, where $U$ is the potential energy relating to the single sub-particle.
 The terms $\nabla U$ and $\nabla Q$ are gradients of the potential energy and of the quantum potential, respectively.
 The third term in the second line describes the viscosity of the medium. As was said above
 the viscosity coefficient in the average is equal to zero.

 Let us rewrite Eq.~(\ref{eq=29}) by regrouping the terms
\begin{equation}
\label{eq=30}
 \nabla\Biggl(
   {{\partial~}\over{\partial\,t}}S + {{1}\over{2m}}(\nabla S)^2 + {{m}\over{2}}{\mathit v}_{_{R}}^2
              + U + Q - \nu(t)\nabla^{2}S
       \Biggr)
 = -m{{\partial~}\over{\partial\,t}}{\vec{\mathit v}}_{_{R}}
   -[{\vec\omega}\times(\nabla S + m{\mathit{\vec v}}_{_{R}})]
    +\nu(t)m\nabla^2 {\vec{\mathit v}}_{_{R}}.
\end{equation}
 We assume that fluctuations of the viscosity about zero occur much more frequent, than characteristic time of displacements of the
 quasi-particles. For that reason, we omit the term $\nu(t)\nabla^{2}S$ by supposing in the first approximation, that the medium is absolutely
 superfluid - there are no energy sources and sinks. By multiplying this equation from the left by ${\vec{\mathit v}}_{_{S}}$ we find that
 the right part of this equation vanishes since 
 $({\vec{\mathit v}}_{_{S}}\cdot{\vec{\mathit v}}_{_{R}})=({\vec{\mathit v}}_{_{S}}\cdot{\vec\omega})=0$.
 On the other part ${\vec{\mathit v}}_{_{S}}\cdot\nabla= {d/d{\mathit l}}$ represents a derivative along the streamline~\citep{KiselevEtAl1999} -
 the left part vanishes along this stream line. The expression under the brackets should be a constant.
 As a result, we come to the following modified Hamilton-Jacobi equation 
\begin{equation}
\label{eq=31}
    {{\partial~}\over{\partial\,t}}S + {{1}\over{2m}}(\nabla S)^2 + {{m}\over{2}}{\mathit v}_{_{R}}^2
      + U({\vec r})
      + {{\hbar^2}\over{2m}}
         \Biggl[
         \Biggl(
            {{\nabla\rho}\over{2\rho}} 
         \Biggr)^2
          - {{\nabla^2\rho}\over{2\rho}}
         \Biggr]
      = C.
\end{equation}
\end{widetext}
 The modification is due to adding the quantum potential~(\ref{eq=25}). In this equation, $C$ is an integration constant. We see that the
 third term in this equation represents energy of the vortex. On the other hand, we can see that the vortex given by Eq.~(\ref{eq=7}) is
 replenished by the kinetic energy coming from the scalar field $S$, namely via the term $({\vec\omega}\cdot\nabla){\vec{\mathit v}}$.
 Solutions of these two equations, Eq.~(\ref{eq=7}) and Eq.~(\ref{eq=31}), describing dynamics of the vortex and scalar fields, depend
 on each other. 

 Both the continuity equation 
\begin{equation}
\label{eq=32}
   {{\partial\,\rho}\over{\partial\,t}} + ({\vec{\mathit v}}\cdot\nabla)\rho=0,
\end{equation}
 which stems from Eq.~(\ref{eq=3}), and the quantum Hamilton-Jacobi equation~(\ref{eq=31}) can be extracted from the following
 Schr\"odinger equation
\begin{equation}
\label{eq=33}
\hspace{-8pt}
  {\bf i}\hbar\,{{\partial\Psi}\over{\partial\,t}}=
  {{1}\over{2m}}(-{\bf i}\hbar\nabla + m{\vec{\mathit v}}_{_{R}})^2\Psi
   +U({\vec r})\Psi - C\Psi.
\end{equation}
 The kinetic momentum operator $(-{\bf i}\hbar\nabla + m{\vec{\mathit v}}_{_{R}})$  contains the term $m{\vec{\mathit v}}_{_{R}}$ describing
 a contribution of the vortex motion. This term is analogous to the vector potential multiplied by the ratio of the charge to the light speed,
 which appears in quantum electrodynamics~\citep{Martins2012}. Appearance of this term in this equation is conditioned by the Helmholtz theorem. 

 By substituting into Eq~(\ref{eq=33}) the wave function $\Psi$ represented in a polar form
\begin{equation}
\label{eq=34}
  \Psi = \sqrt{\,\rho\,}\exp\{{\bf i}S/\hbar\}
\end{equation}
 and separating on real and imaginary parts we come to Eqs.~(\ref{eq=31}) and~(\ref{eq=32}). So, the Navier-Stokes equation~(\ref{eq=6})
 with the slightly expanded the pressure gradient term can be reduced to the Schr\"odinger equation if we take into consideration also the
 continuity equation.

 The Shr\"odinger wave equation can be resolved by heuristic writting of a solution through using the Huygens' principle in its mathematical
 integral presentation
\begin{equation}
  \Psi(x,t_2) = \int G(x,y) \Psi(y,t_1)dy, ~~~ t_2>t_1.
\label{eq=35}
\end{equation}
 The function $G(x,y)$ is called a propagator. 
\begin{figure*}[htb!]
  \centering
  \begin{picture}(200,430)(130,18)
      \includegraphics[scale=0.75]{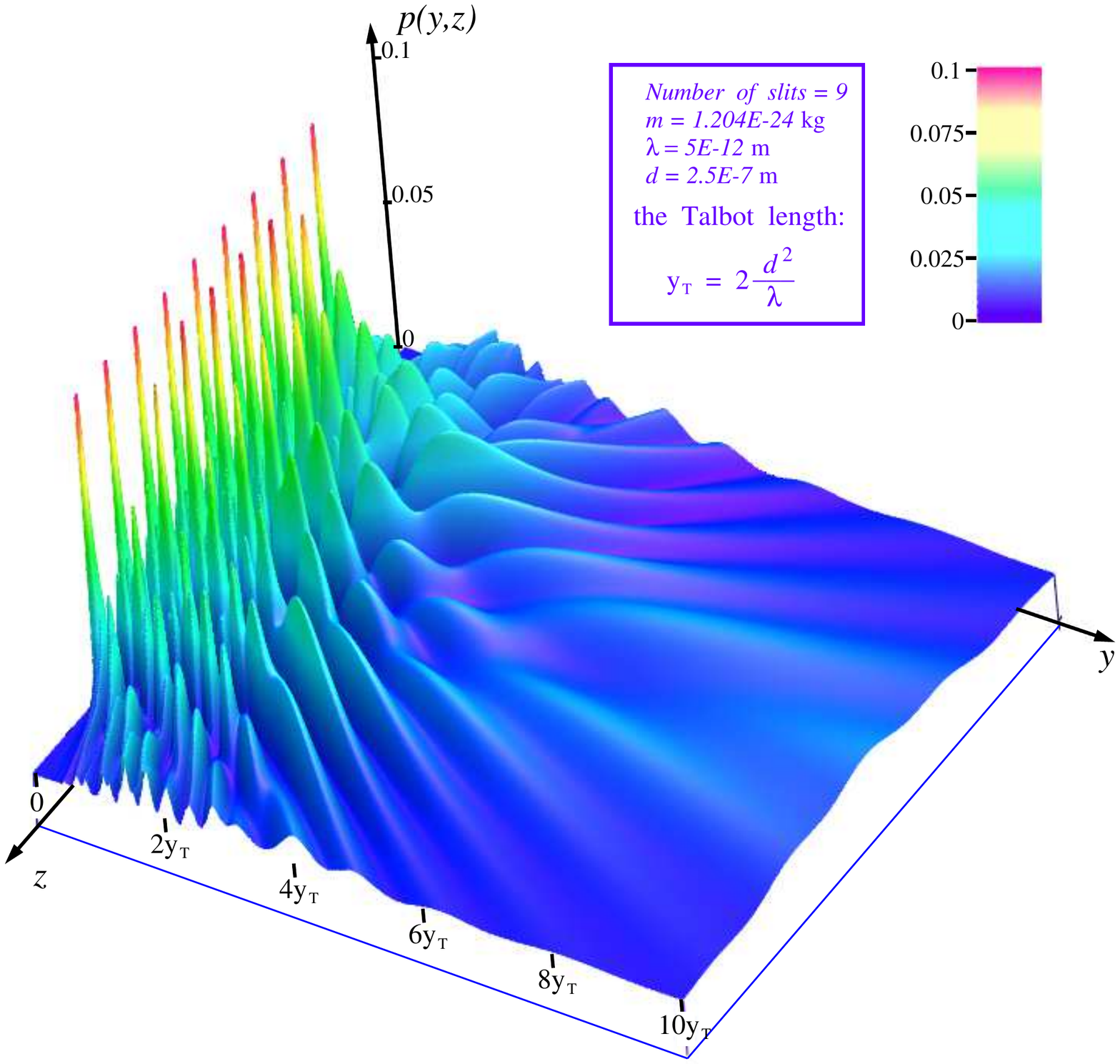}
  \end{picture}
  \caption{
  Probability density distribution from scattering the fullerene molecules (60 carbon atoms packed on the sphere like a football)
  on the grating containing 9 slits: de Broglie wavelength is 5 pm and the distance between slits is 250 nm.
  }
  \label{fig=7}
\end{figure*}
\begin{figure*}[htb!]
  \centering
  \begin{picture}(200,380)(160,10)
      \includegraphics[scale=0.65]{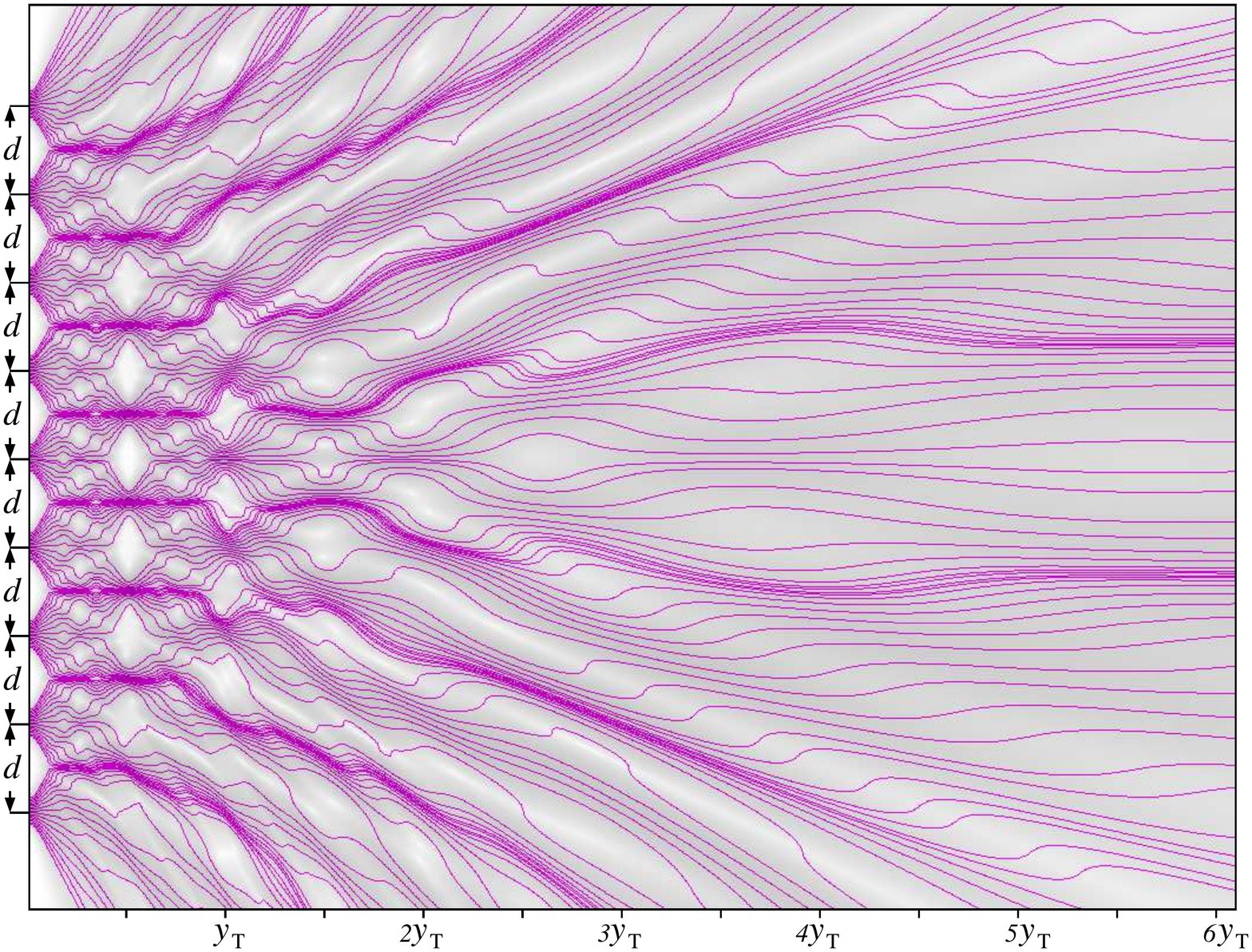}
  \end{picture}
  \caption{
  Interference pattern of the coherent flow of the fullerene molecules with the de Broglie wavelength $\lambda=5$ pm within
  a zone $y\le 6y_{_{\bf T}}$ from the grating containing 9 slits. Lilac curves against the grey background represent the Bohmian trajectories.
  }
  \label{fig=8}
\end{figure*}
 It bears information about a physical scene containing physical equipment such as sources, detectors, collimators, gratings, etc.
 This scene is simulated by the potential $U({\vec r})$ embedded in the Schr\"odinger equation and by the boundary conditions.
 So, a solution of the Schr\"odinger equation can be achieved by applying the Feynman path integral technique~\citep{FeynmanHibbs1965, Derbes1996}.
 By applying this technique for describing the wave propagation through a lattice consisting of $N$ slits
 we get the solution~\citep{Sbitnev2010a}
\begin{widetext}
\begin{equation}
 |\Psi(y,z)\rangle =
 {{1}\over{N\sqrt{\displaystyle 1 + {\bf i} {{\lambda y}\over{2\pi b^{2}}} }}} 
 \cdot
   \sum\limits_{n=0}^{N-1}
  \exp
  \left\{
   \matrix{
   - {{\displaystyle \Biggl(z - \Biggl(n - {{N-1}\over{2}}
                                \Biggr)d
                     \Biggr)^{2}}\over{\displaystyle 2b^2\Biggr(
                                                                 1 + {\bf i} {{\lambda y}\over{2\pi b^{2}}}
                                                         \Biggl)}}
          }
  \right\}
\label{eq=36}
\end{equation}
\end{widetext}
 Here  ${\bf i} = \sqrt{-1}$ is imaginary unit, $\lambda$ is the de Broglie wavelength, $b$ is the slit width,
 and  $d$ is the distance between slits.
 In this calculation we have used $\lambda = 5$~pm, $b=5\cdot10^{3}\lambda$,
 and $d=5\cdot10^{4}\lambda = 10b = 250$ nm.
 By choosing $N = 9$ slits, for example, we find the interference pattern shown in Fig.~\ref{fig=7}
 as the density distribution function~\citep{Sbitnev2010a}. 
 This function is a scalar product of the wave function $|\Psi(y,z)\rangle$, namely:
\begin{equation}
\label{eq=37}
 p(y,z) = \langle\Psi(y,z)|\Psi(y,z)\rangle.
\end{equation}
 A useful unit of length at observation of the interference patterns is the Talbot length: 
\begin{equation}
     y_{_{\bf T}} = 2\,{{d^2}\over{\lambda}}.
\label{q=38}
\end{equation}
 This length bears name of Henry Fox Talbot who discovered in 1836~\citep{Talbot1836} a beautiful interference pattern of light,
 named further as the Talbot carpet~\citep{BerryKlein1996, BerryEtAl2001}.

 The particles, incident on the slit grating, come from a distant coherent source. The de Broglie wavelength of the particle, $\lambda=h/p$ 
 ($h$ is the Planck constant, and $p$ is the particle momentum) is a main characteristics binding the corpuscular Newtonian physics with the
 wave Huygens' physics. It is that we call now the wave-particle dualism.  The de Broglie pilot wave being represented by the complex-valued
 wave function $|\Psi\rangle$  fills all ambient space, except of opaque objects, which determine the boundary conditions. The particle passes
 from the source to a place of detection along the optimal trajectory, Bohmian trajectory~\citep{Sbitnev2012}.
 The equation describing motion of the Bohmian particle can be found, for example, in~\citep{BensenyEtAl2014}.
 There is the unique trajectory for each the particle, the vortex ball in our case. However, an attempt to measure exact position of the ball
 along the trajectory together with its velocity fails. Namely, there is no way to measure simultaneously the complementary parameters,
 such as coordinate and velocity, what follows from the uncertainly principle~\citep{Sbitnev2013a}. 

 Figure~\ref{fig=8} shows in lilac color Bohmian trajectories divergent from the slit grating. The probability density distribution is shown
 here in grey color ranging from white for $p = 0$ to light grey for $\max p$.  Bundle of the Bohmian trajectories imitates a fluid flow
 through the obstacle, containing slits, relatively well. One can see that characteristic streamlets are formed in the flow, along which
 particles move.  Such a vision of hydrodynamical behaviors of quantum systems is typical for many scientists since the formation of the
 quantum mechanics up to our
 days~\citep{Madelung1926, BohmVigier1954, Grossing2004, Wyatt2005, JungelMili2012, VanFulop2006, DamazioSiqueira2013}. 
 Principal moment is that the Schr\"odinger equation describes the expiration of the superfluid
 medium, which depend on the boundary conditions and other devices perturbing it (as, for example, the slit gratings, collimators and others).  The vortex balls move along optimal directions of the flows - along the Bohmian paths.  

\section{\label{sec:level4}Physical vacuum as a superfluid medium}

 The Schr\"odinger equation~(\ref{eq=33}) describes a flow of the peculiar fluid that is the physical vacuum. The vacuum contains pairs of
 particle-antiparticles. The pair, in itself, is the Bose particle that stays at a temperature close to zero. In aggregate, the pairs make up
 Bose-Einstein condensate.  It means that the vacuum represents a superfluid medium~\citep{SinhaEtAl1976}.
 A 'fluidic' nature of the space itself is exhibited through this medium.
 Another name of such an 'ideal fluid' is the ether~\citep{Martins2012}.
\begin{figure}
  \centering
  \begin{picture}(200,150)(20,35)
      \includegraphics[scale=0.6]{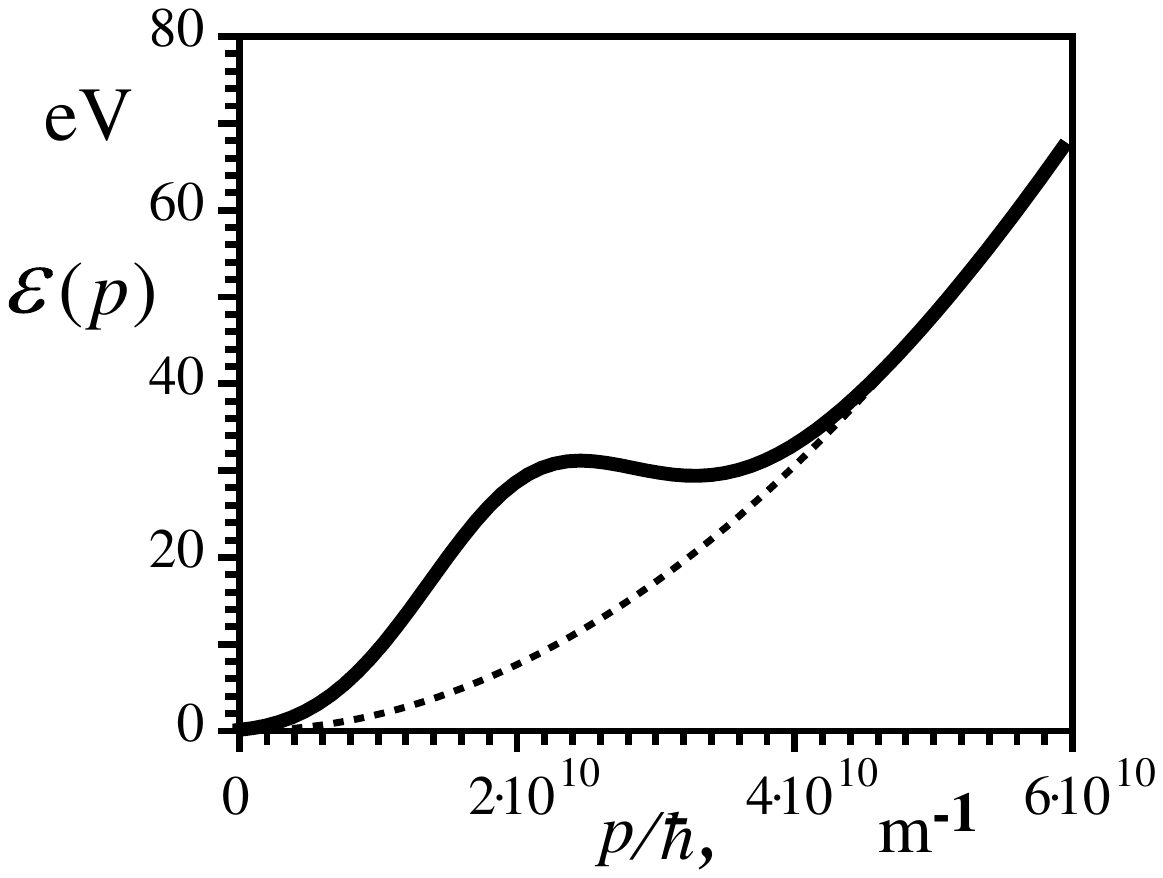}
  \end{picture}
  \caption{
  The dispersion relation $\epsilon$ vs. $p$. The dotted curve shows the non-relativistic square dispersion relation
  $\epsilon\sim p^{2}$ . The hump on the curve is a contribution of the roton component
   $p_{_{R}}f(p-p_{_{R}})$, $p_{_{R}}/\hbar\approx 1.89\cdot10^{10}$ m$^{-1}$, and $\sigma=0.5p_{_{R}}$.
  }
  \label{fig=9}
\end{figure}

 The physical vacuum is a strongly correlated system with dominating collective effects~\citep{Roberts2001}
 and the viscosity equal to zero. Nearest analogue of such a medium is the superfluid helium~\citep{Volovik2003},
 which will serve us as an example for further consideration of this medium. The vacuum is defined as a state with the lowest possible energy.
 We shall consider a simple vacuum consisting of electron-positron pairs. The pairs fluctuate within the first Bohr orbit having energy about
 $13.6\cdot2$ eV $\approx 27$ eV. Bohr radius of this orbit is $r_{\,1}\approx5.29\cdot10^{-11}$ m. These fluctuations occur about the center
 of their masses. The total mass of the pair, $m_p$, is equal to doubled mass of the electron, $m$. The charge of the pair is zero. The total
 spin of the pair is equal to 0. The angular momentum, $L$, is nonzero, however. For the first Bohr orbit $L=\hbar$. The velocity of rotating
 about this orbit is $L/(r_{\,1}m)\approx 2.192\cdot10^{\,6}$  m/s. It means that there exist an elementary vortex. Ensemble of such vortices
 forms a vortex line.
\begin{figure}
  \centering
  \begin{picture}(200,260)(0,10)
      \includegraphics[scale=0.6]{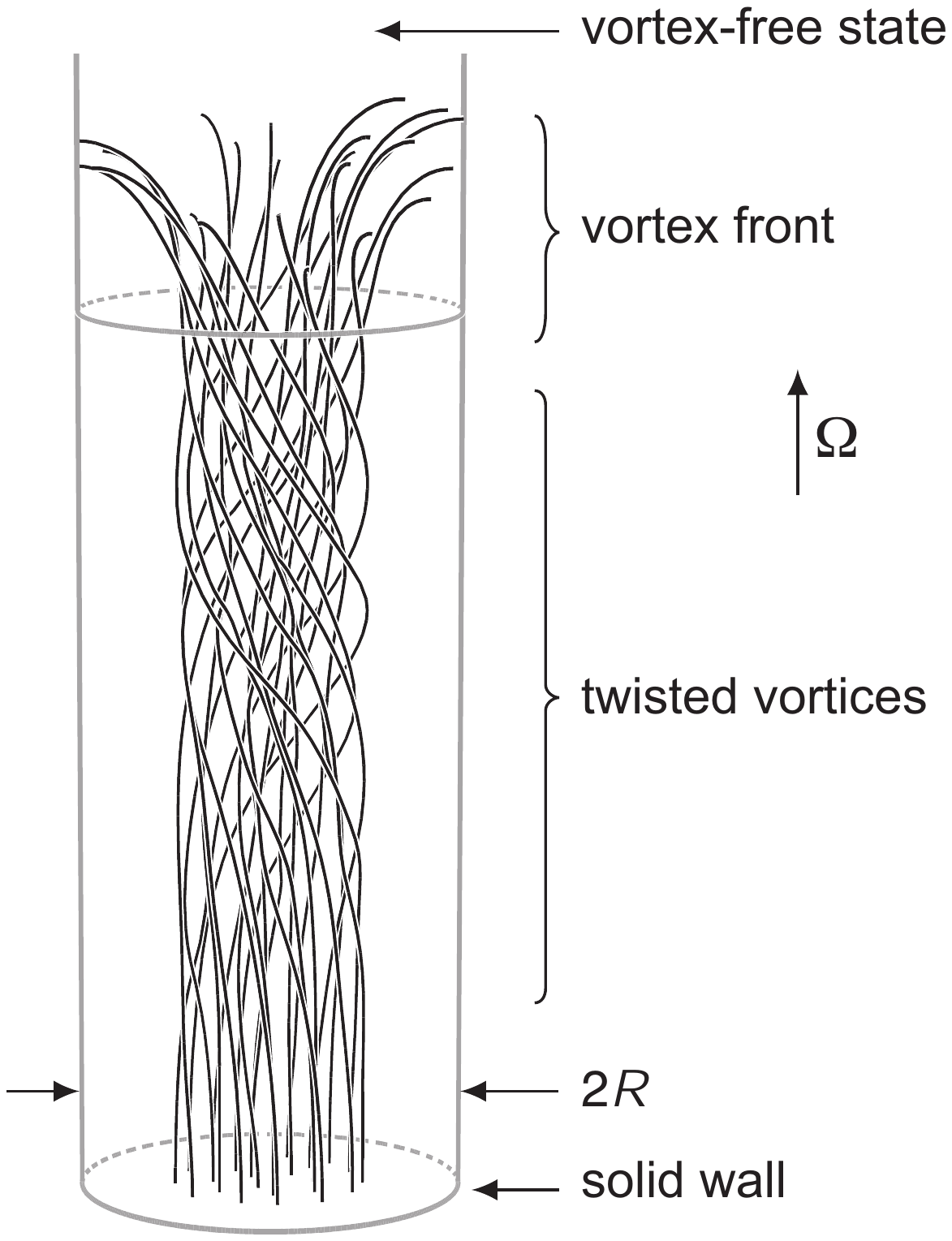}
  \end{picture}
  \caption{
 The formation of twisted vortex state~\citep{EltsovEtAl2006}. The vortices have their propagating ends bent to the side wall of the rotating cylinder.
 As they expand upwards into the vortex-free state, the ends of the vortex lines rotate around the cylinder axis.
 The twist is nonuniform because boundary conditions allow it to unwind at the bottom solid wall. The figure gives a snapshot
 (at time $t=25\Omega^{-1}$, where $\Omega$  is the angular velocity.) of a numerical simulation of 23 vortices initially generated near
 the bottom end ($t = 0$). Courtesy kindly by Erkki Thuneberg.
  }
  \label{fig=10}
\end{figure}
\begin{figure}
  \centering
  \begin{picture}(200,200)(35,30)
      \includegraphics[scale=0.45]{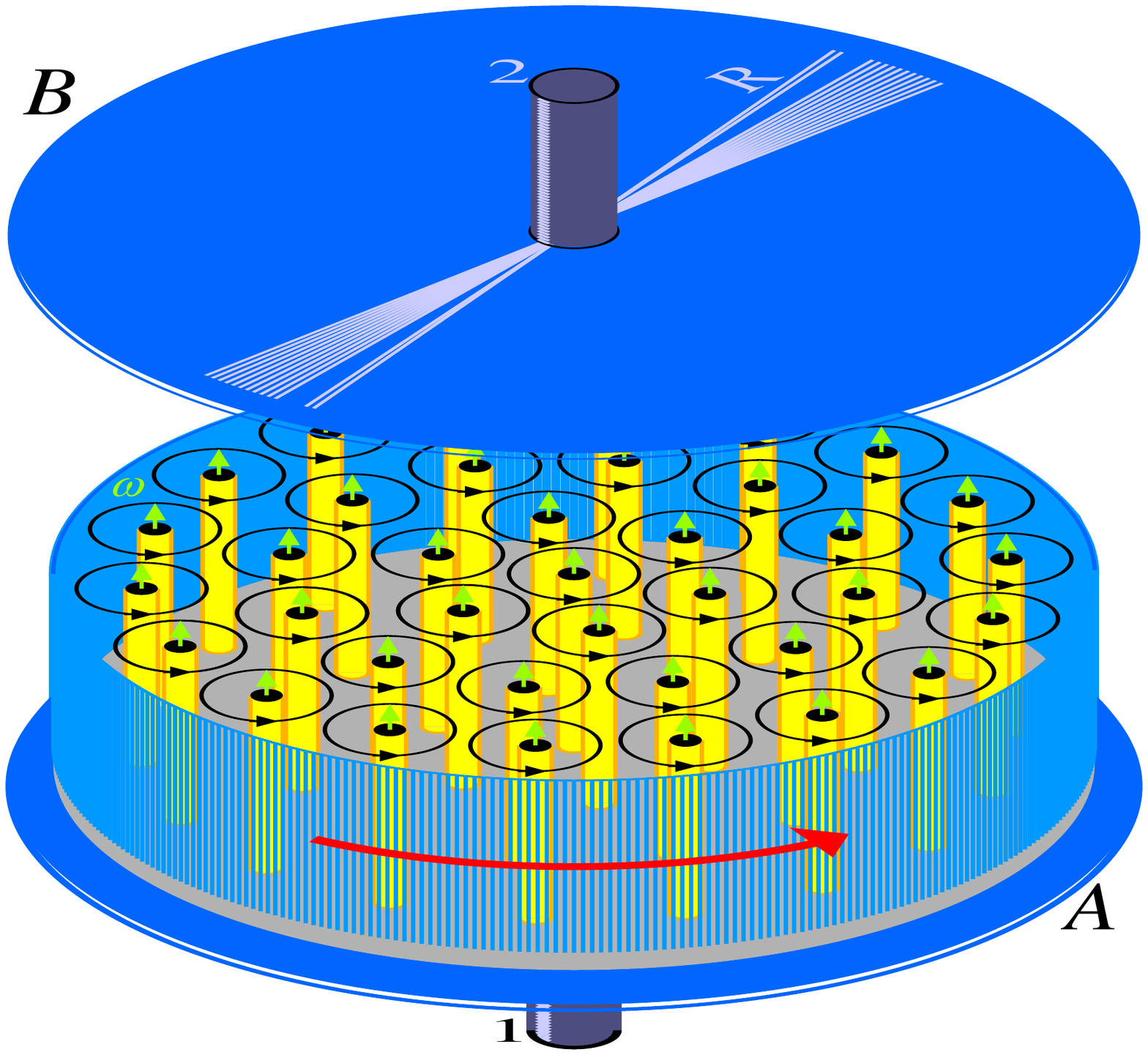}
  \end{picture}
  \caption{
  Rotation of the superfluid fluid is not uniform but takes place via a lattice of quantized vortices, whose cores (colored in yellow)
 are parallel to the axis of rotation~\citep{EltsovEtAl2006, LounasmaaAndThuneberg1999}. Green arrows are the vorticity $\omega$. Small black
 arrows indicate the circulation of the velocity ${\mathit v}_{_{R}}$ around the cores. The vacuum is supported between two non-ferromagnetic
 disks, {\it A} and {\it B}, fixed on center shafts, {\bf 1} and {\bf 2}, of electric motors~\citep{Samohvalov2013}, see Fig. 12.
 Radiuses of the both disks are $R = 82.5$ mm and distance between them can vary from 1 to 3 mm and more. The vortex bundle rotates rigidly
 with the disk {\it A}. As soon as the vortex bundle reaches the top disk {\it B} it begins rotation.
  }
  \label{fig=11}
\end{figure}
\begin{figure*}
  \centering
  \begin{picture}(200,170)(180,60)
      \includegraphics[scale=1.2]{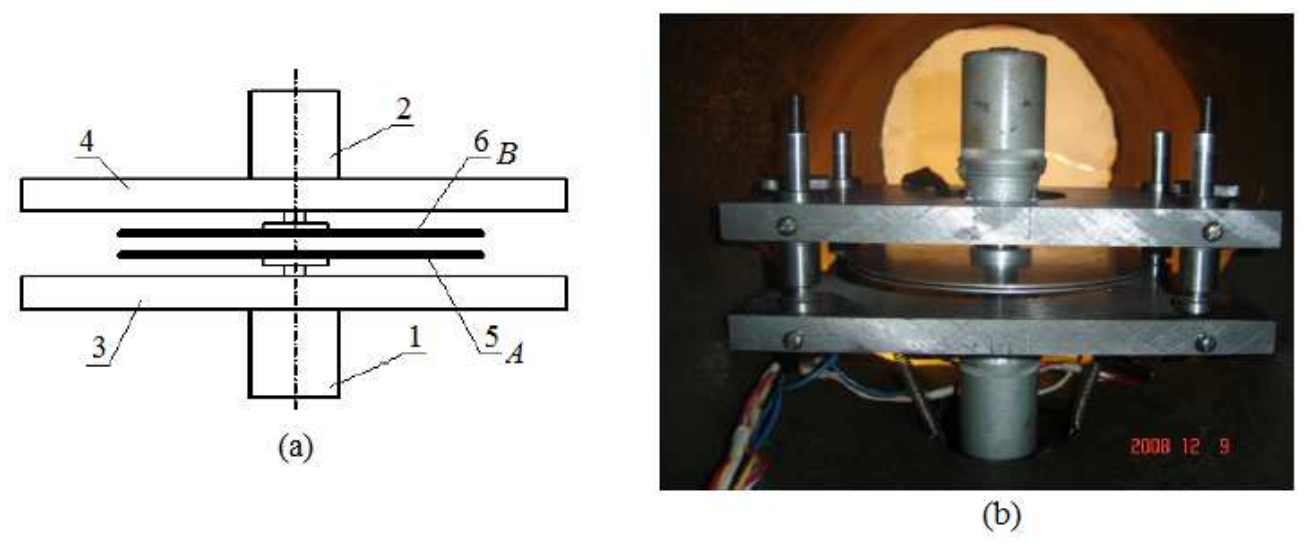}
  \end{picture}
  \caption{
   Basic diagram (a) and general view of the device (b) for researching mass dynamics effects~\citep{Samohvalov2013}: 1 and 2 are shafts with
 mounted on them electric motors; 3 and 4 are steel plates with mounted on them electromagnetic brakes; 5 and 6
 ({\it A} and {\it B}, see Fig.~\ref{fig=11}) are disks rigidly fixed on flanges of the rotors of the electric motors.
 Courtesy kindly by Vladimir Samokhvalov.
  }
  \label{fig=12}
\end{figure*}

 We may evaluate the dispersion relation between the energy, $\epsilon(p) =\hbar\omega$, and wave number,
 $p=\hbar k$,
 as it done in~\citep{LeGal2013}.
 As follows from the Schr\"odinger equation~(\ref{eq=33}) we have:
\begin{equation}
\label{eq=39}
 \epsilon(p) = {{1}\over{2m_{p}}}
   (p + p_{_{R}}f(p-p_{_{R}}) )^{2}.
\end{equation}
 Here $p_{_{R}} = L/r_{\,1}=m_{p}{\mathit v}_{_{R}}$ is the momentum of the rotation. The function $f(p - p_{_{R}})$ is a form-factor
 relating to the electron-positron pairs rotating about the center of their mass $m_{p}$. The form-factor describes dispersion of the momentum
 $p$ around $p_{_{R}}$ conditioned by fluctuations about the ground state with the lowest energy.
 The form-factor is similar to the Gaussian curve
\begin{equation}
\label{eq=40}
  f(p-p_{_{R}}) = \exp\Biggl\{-
                              {{(p-p_{_{R}})^2}\over{2\sigma^2}}
                      \Biggr\}.
\end{equation}
 Here $\sigma$ is the variance of this form-factor. It is smaller or close to $p_{_{R}}$.
 The dispersion relation~(\ref{eq=39}) is shown in Fig.~\ref{fig=9}. The hump on the curve is due to the contribution of the rotating
 electron-positron pair about the center of their masses. These rotating objects are named rotons~\citep{LeGal2013}. 

 Rotons are ubiquitous in vacuum because of a huge availability of pairs of particle-antiparticle.
 The movement of the roton in the free space is described by the Schr\"odinger equation
\begin{equation}
\label{eq=41}
  {\bf i}\,{{\partial\Psi}\over{\partial\,t}}=
  {{1}\over{2m_{p}}}(-{\bf i}\hbar\nabla + m_{p}{\vec{\mathit v}}_{_{R}})^2\Psi
   - C\Psi.
\end{equation}
 The constant $C$ determines an uncertain phase shift of the wave function, and most possible this phase relates to the chemical potential
 of a boson (the electron-positron pair)~\citep{LeGal2013}. We shall not take into account contribution of this term in the dispersion diagram
 because of its smallness. As follows from the above consideration of the Navier-Stokes equation, Eq.~(\ref{eq=41}) can be reduced to the
 Euler equation
\begin{equation}
\label{eq=42}
 {{\partial {\vec{\mathit v}}_{_{R}} }\over{\partial\,t}} + [{\vec\omega}\times{\vec{\mathit v}}_{_{R}}]
  = - {{\nabla P}\over{m_{p}\,\rho}},
\end{equation}
 that describes a flow of the inviscid incompressible fluid under the pressure field $P$. One can see from here that the Coriolis force appears
 as a restoring force, forcing the displaced fluid particles to move in circles. The Coriolis force is the generating force of waves called
 inertial waves~\citep{LeGal2013}. The Euler equation admits a stationary solution for uniform swirling flow
 under the pressure gradient along $z$. 

 Formation of the swirling flow, the twisted vortex~ state, has been studied in the superfluid~ $^3$He-B~\citep{EltsovEtAl2006}.
 These observations give us a possibility to suppose the existence of such phenomena in the physical vacuum. The twisted vortex states observed
 in the superfluid $^3$He-B are closely related to the inertial waves in rotating classical fluids. The superfluid initially is at
 rest~\citep{EltsovEtAl2006}. The vortices are nucleated at a bottom disk platform rotating with the angular velocity $\Omega$ about axis~$z$.
 As the platform rotates they propagate upward by creating the twisted vortex state spontaneously, Fig.~\ref{fig=10}. The Coriolis forces take
 part in this twisting. The twisted vortices grow upward along the cylinder axis~\citep{LounasmaaAndThuneberg1999}. 

 Analogous experiment with nucleating vortices can be realized when the lower disk A rotates in the vacuum, Fig.~\ref{fig=11}. In this case,
 the vortices are viewed as the dancing electron-positron pairs on the first Bohr orbit. As the vortices grow upward the spontaneous twisted
 vortex states arise. The latter by reaching upper fixed disk {\it B} can capture it into rotation.

 Pr. V. Samohvalov has shown through the experiment~\citep{Samohvalov2013}, that the vortex bundle induced by rotating the bottom
 non-ferromagnetic disk {\it A} leads to rotation of the upper fixed initially non-ferromagnetic disk {\it B}, Fig.~\ref{fig=12}. Both disks
 at room temperature have been placed in the container with technical vacuum at 0.02 Torr, The utmost number of the vortices that may be placed
 on the square of the disk {\it A} is $N_{\max}  = (2\pi R^2)/(2\pi r_{\,1}^2) = 2\cdot10^{18}$, where $R = 82.5$ mm is the radius of the disk
 and $r_{\,1}\approx 5.29\cdot10^{-11}$  m is the radius of the first Bohr orbit. Really, the number of the vortices situated on the square,
 $N$, is considerably smaller. It can be evaluated by multiplying $N_{\max}$ by a factor $\delta$. This factor is equal to the ratio of the
 geometric mean of the velocities ${\mathit v}_{_{R}}=\hbar/(r_{\,1}m)\approx 2.192\cdot10^{6}$ m/s and $V_{_{D}}=R\Omega$ to their arithmetic
 mean. Here $\Omega$ is  an angular rate of the disk {\it A}. So, we have
\begin{equation}
\label{eq=43}
\hspace{-12pt}
 N = N_{\max} {{ \sqrt{ {\mathit v}_{_{R}} \cdot V_{_{D}} } }\over{ {\mathit v}_{_{R}}+ V_{_{D}} }}
   = N_{\max} \sqrt{ {{V_{_{D}}}\over{{\mathit v}_{_{R}}}} } \approx 6\cdot10^{15}
\end{equation}
 at the angular rate $\Omega  = 160$ 1/s~\citep{Samohvalov2013} the disk velocity $V_{_{D}} = 13.2$ m/s. Now we can evaluate the kinetic energy
 of the vortex bundle induced by the rotating disk {\it A}.This kinetic energy is $E = N\cdot m_{p}{\mathit v}_{_{R}}^2/2 \approx 0.026$ J.
 This energy is sufficient for transfer of the moment of force to the disk {\it B}. Measured in the experiment~\citep{Samohvalov2013}
 the torque is about 0.01 N$\cdot$m. So, the disk B can be captured by the twisted vortex. 

 The formation of the growing twisted vortices can be confirmed with attracting modern methods of interference of light rays passing through
 the gap between the disks. Light traveling along two paths through the space between the disks undergoes a phase shift manifested in the
 interference pattern~\citep{Martins2012, BerryEtAl1980} as it was shown in the famous experiment of~\citet{AharonovBohm1959}.

\section{\label{sec:level5}Conclusion}

 The Schr\"odinger equation is deduced from two equations, the continuity equation and the Navier-Stokes equation. At that, the latter contains
 slightly modified the gradient pressure term, namely, $\nabla P\rightarrow \rho_{_{M}}\nabla(P/\rho_{_{M}}) - P\nabla\ln(\rho_{_{M}})$.
 The extra term $P\nabla\ln(\rho_{_{M}})$  describes change of the pressure induced by change of the entropy $\ln(\rho_{_{M}})$  per length.
 In this case, the modified gradient pressure term can be reduced to the quantum potential through using the Fick's law.
 In the law we replace also the diffusion coefficient by the factor $\hbar/2m$, where $\hbar$ is the reduced Planck constant
 and $m$ is mass of the particle. 

 We have shown that a vortex arising in a fluid can exist infinitely long if the viscosity undergoes periodic oscillations between positive
 and negative values. At that, the viscosity, in average on time, stays equal to zero. It can mean that the fluid is superfluid. In our case,
 the superfluid consists of pairs of particle-antiparticle representing the Bose-Einstein condensate.

 As for the quantum reality, such a periodic regime can be interpreted as exchange of the energy quanta of the vortex with the vacuum through
 the zero-point vacuum fluctuations. In reality, these fluctuations are random, covering a wide range of frequencies from zero to infinity.
 Based on this observation we have assumed that the fluctuations of the vacuum ground state can support long-lived existence of vortex quantum
 objects. The core of such a vortex has nonzero radius inside of which the velocity tends to zero. In the center of the vortex, the velocity
 vanishes. The velocity reaches maximal values on boundary of the core, and then it decreases to zero as the distance to the vortex goes to
 infinity. 

 The experimental observations of the Couder's team~\citep{CouderForte2006, CouderEtAll2005, ProtiereEtAll2006, EddiEtAll2011}
 can have far-reaching ontological perspectives in regard of studying our universe. Really, we can imagine that our world is represented by
 myriad of baryonic and lepton "droplets" bouncing on a super-surface of some unknown dark matter. A layer that divides these "droplets",
 i.e., particles, and the dark matter is the superfluid vacuum medium. This medium, called also the ether~\citep{Nelson1966},
 is populated by the particles of matter ("droplets"), which exist in it and move
 through it~\citep{Martins2012, PinheiroBuker2012, SaraveniEtAl2013}.
 The particle traveling through this medium perturbs virtual particle-antiparticle pairs, which, in turn, create both constructive and
 destructive interference at the forefront of the particle~\citep{FeynmanHibbs1965}. Thus, the virtual pairs interfering each other provide an optimal, Bohmian,
 path for the particle.

 Assume next, that the baryonic matter is similar, say, on "hydrophobic" fluid, whereas the dark matter, say, is similar to "hydrophilic" fluid.
 Then the baryonic matter will diverge each from other on cosmological scale owing to repulsive properties of the dark matter, like soap spots
 diverge on the water surface. Observe that this phenomenon exhibits itself through existence of the short-range repulsive gravitational force
 that maintains the incompatibility between the dark matter and the baryonic matter~\citep{FamaeyMcGaugh2012, Chung2014}.
 At that, the dark matter stays invisible. One can imagine that the zero-point vacuum fluctuations are nothing as weak ripples on a surface of
 the dark matter.

\begin{acknowledgments}

 The author thanks Mike Cavedon for useful and valuable remarks and offers.
 The author thanks also Miss Pipa (Quantum Portal administrator) for preparing a program drawing Fig.~\ref{fig=8}.

\end{acknowledgments}

\section*{\label{sec:level6}
 Appendix: Nelson's derivation of the Schr\"odinger equation
}

 Nelson proclaim that the medium through which a particle moves contains myriad sub-particles that accomplish Brownian motions by colliding
 with each other chaotically. The Brownian motions is described by the Wiener process with the diffusion coefficient
\begin{equation}
\label{eq=44}
  {\bar\nu} = {{\hbar}\over{2m}}.
\end{equation}
 Here $m$ is mass of the particle and $\hbar=h/2\pi$  is the reduced Planck constant.
 Here we use $\nu$ with the upper bar in order to avoid confusion with the kinematic viscosity adopted in hydrodynamics.
 As seen this motion has a quantum nature~\citep{Nelson1966} in contrast to the macroscopic Brownian motions where the diffusion coefficient
 has a view ${\bar\nu}=kT/m\beta^{-1}$; here $k$ is Boltzmann constant, $T$ is a temperature, and  is the relaxation time.

 Two equations are main in the article~\citep{Nelson1966}. The position ${\bf x} (t )$ of the Brownian particle, being subjected
 either by external forces or by currents in the medium, can be written by two equivalent equations
\begin{eqnarray}
\label{eq=45}&&
  d{\bf x}(t)={\bf b}({\bf x}(t),t)dt~ + d{\bf w}(t),\\
\label{eq=46}&&
  d{\bf x}(t)={\bf b}_{*}({\bf x}(t),t)dt + d{\bf w}_{*}(t).
\end{eqnarray}
 Here  ${\bf w}(t)$ and ${\bf w}_{*}(t)$  are the Wiener processes, both have equivalent properties.
 Variables ${\bf b}$ and ${\bf b}_{*}$  are vector-valued forward and backward functions on space-time, respectively.
 In fact, they are the mean forward and mean backward measured quantities
\begin{eqnarray}
\label{eq=47}\hspace{-22pt}
  {\bf b}({\bf x}(t),t )~ &=& \lim\limits_{\Delta t\rightarrow 0_{+}}
   E_{\,t}{{{\bf x}(t+\Delta t)-{\bf x}(t)}\over{\Delta t}}, \\
\label{eq=48}\hspace{-22pt}
  {\bf b}_{*}({\bf x}(t),t ) &=& \lim\limits_{\Delta t\rightarrow 0_{+}}
   E_{\,t}{{{\bf x}(t)-{\bf x}(t-\Delta t)}\over{\Delta t}}.
\end{eqnarray}
 Here $E_{\,t}$ denotes the conditional expectation (average) given the state of the system at time $t$, and $0_{+}$ means
 that $\Delta t$  tends to $0$ through positive values.
 Thus ${\bf b}({\bf x}(t),t)$  and ${\bf b}_{*}({\bf x}(t),t)$ are again stochastic variables~\citep{Nelson1967, Nelson1985}.
 It is instructive to compare calculus~(\ref{eq=47}) and~(\ref{eq=48}) with classical calculations of infinitesimal small increments
\begin{eqnarray}
\nonumber
  {\mathit v}(t) &=& \lim\limits_{\Delta t\rightarrow 0_{+}} {{x(t+\Delta t)-x(t)}\over{\Delta t}} \\
                 &=& \lim\limits_{\Delta t\rightarrow 0_{+}} {{x(t)-x(t-\Delta t)}\over{\Delta t}}.
\label{eq=49}
\end{eqnarray}
 These calculations are seen to be symmetrical with respect to the time arrow,
 whereas~(\ref{eq=47}) and~(\ref{eq=48}) are not.

 It should be noted that ${\bf b}({\bf x}(t),t)$  and ${\bf b}_{*}({\bf x}(t),t)$ are not real velocities. The real current velocity of
 the particle is calculated as
\begin{equation}
\label{eq=50}
  {\vec{\mathit v}}(t) = {{1}\over{2}}\biggl(
                                             {\bf b}({\bf x}(t),t) + {\bf b}_{*}({\bf x}(t),t)
                                      \biggr).
\end{equation}
 There is a one more velocity, which is represented via difference of ${\bf b}({\bf x}(t),t)$  and ${\bf b}_{*}({\bf x}(t),t)$:
\begin{equation}
\label{eq=51}
  {\vec{u}}(t) = {{1}\over{2}}\biggl(
                                     {\bf b}({\bf x}(t),t) - {\bf b}_{*}({\bf x}(t),t)
                              \biggr).
\end{equation}
 According to Einstein's theory of Brownian motion, ${\vec{u}}(t)$ is the velocity acquired by a Brownian particle, in equilibrium
 with respect to an external force, to balance the osmotic force~\citep{Nelson1967}.
 For this reason, this velocity is named the osmotic velocity. It can be expressed in the following form
\begin{equation}
\label{eq=51}
   {\vec{u}}(t) = {\bar\nu}\nabla(\ln(\rho(t)) = {{\hbar}\over{m}}{{\nabla  R(t)}\over{R(t)}},
\end{equation}
 where $\rho(t)$ is the probability density of disclosing a particle in the vicinity of ${\bf x}(t)$
 and $R(t) =\rho(t)^{1/2}$ is the probability density amplitude.
 In turn, the current velocity is expressed through gradient of a scalar
 field $S$ called the action
\begin{equation}
\label{eq=52}
  {\vec{\mathit v}} = -{{\hbar}\over{m}}\nabla S(t).
\end{equation}

 The both equations, Eqs.~(\ref{eq=45}) and~(\ref{eq=46}), introduced above are important for derivation of the Schr\"odinger equation.
 The derivation of the equation is provided by the use of the wave function presented in the polar form
\begin{equation}
\label{eq=53}
  \Psi = R\exp\{{\bf i}S/\hbar\},
\end{equation}
 by replacing the velocities ${\vec{\mathit v}}(t)$  and ${\vec u}(t)$  in the initial equations. It should be noted that Nelson departs
 from two equations describing directed the forward and backward Brownian motions which are written down for real-valued functions.
 In order to come to the Schr\"odinger equation he has used a complex-valued wave function $\exp\{R+{\bf i}S\}$ instead of the generally
 accepted $R\exp\{{\bf i}S/\hbar\}$. Obviously, this discrepancy are eliminated by replacing $\exp\{R\}\rightarrow R$.

 Observe that the wave function represented in the polar form~(\ref{eq=53}) is used for getting equations underlying
 the Bohmian mechanics~\citep{BensenyEtAl2014}. These two equations are the continuity equation and the Hamilton-Jacobi equation containing
 an extra term known as the Bohmian quantum potential. The quantum potential has the following view:
\begin{eqnarray}
\nonumber
  Q &=& -{{\hbar^2}\over{2m}}{{\nabla^2 R}\over{R}} = -{{\hbar}\over{2}}{{(\nabla R{\vec u})}\over{R}}\\
    &=& -{{m}\over{2}}u^2 - {{\hbar}\over{2}}(\nabla {\vec u}).
\label{eq=54}
\end{eqnarray}
 One can see that the quantum potential depends only on the osmotic velocity, which is expressed through difference of the forward and backward
 averaged quantities~(\ref{eq=47}) and~(\ref{eq=48}). These forward and backward quantities can be interpreted as uncompensated flows through
 a 'semipermeable membrane' which represents an instant dividing the past and the future. Following to~\citet{LicataFiscaletti2014}, who have shown
 that the quantum potential has relation to the Bell length indicating a non-local correlation, one can add that the non-local correlation
 exists also between the past and the future. E. Nelson as one can see has considered a particle motion through the ether populated by
 sub-particles experiencing accidental collisions with each other. The Brownian motions of the sub-particles submits to the Wiener process
 with the diffusion coefficient $\nu$ proportional to the Plank constant as shown in Eq.~(\ref{eq=44}).
 The ether behaves itself as a free-friction fluid.




\end{document}